\documentclass[aps,reprint,amsmath,amssymb,groupaddress,superscriptaddress,pra]{revtex4-2}
\usepackage{newtxtext}
\usepackage{newtxmath}
\usepackage{graphicx}
\usepackage[utf8]{inputenc}
\usepackage{amsmath}
\usepackage{subfigure}
\usepackage{amsfonts}
\usepackage{amssymb}
\usepackage{bm}
\usepackage{dcolumn}
\usepackage{color}
\usepackage[T1]{fontenc}
\usepackage{booktabs, array, float, tabularx, lipsum, multirow, mathtools}
\graphicspath{{figs/}{figsgaoerb/}} 
\usepackage{soul}
\usepackage{CJK}
\usepackage{xcolor}
\usepackage{booktabs}

\usepackage[colorlinks,citecolor=blue,linkcolor=blue,anchorcolor=blue,urlcolor=blue]{hyperref}

\begin{document}
\begin{CJK*}{UTF8}{gbsn}

\title{Macroscopic Thermodynamic Framework for the Mpemba Effect}

\author{Yun-Qian Lin (林蕴芊)}
\affiliation{School of Physics and Astronomy, Beijing Normal University, Beijing, 100875, China}

\author{Z. C. Tu (涂展春)}
\address{School of Physics and Astronomy, Beijing Normal University, Beijing, 100875, China}
\address{Key Laboratory of Multiscale Spin Physics (Beijing Normal University), Ministry of Education, Beijing 100875, China}

\author{Yu-Han Ma (马宇翰)}
\email{yhma@bnu.edu.cn}
\address{School of Physics and Astronomy, Beijing Normal University, Beijing, 100875, China}
\address{Key Laboratory of Multiscale Spin Physics (Beijing Normal University), Ministry of Education, Beijing 100875, China}

\date{\today}

\begin{abstract}
The counterintuitive Mpemba effect, wherein a hotter system cools faster, critically lacks a general macroscopic theory. Here, starting from linear irreversible thermodynamics, we formulate a generalized Newton's cooling law, $\mathrm{d}T/\mathrm{d}t = -[\gamma_0 + \mathcal{M}Q(t)](T-T_r)$, for a system at temperature $T$ relaxing in a thermal reservoir at $T_r$, where the bare relaxation rate $\gamma_0$ is modified by an initial-state memory term, $Q(t) \propto T(0)-T_r$. Arising from the interplay between heat flux and structural evolution, the coefficient $\mathcal{M}$ governs anomalous relaxation behaviors, where $\mathcal{M} > 0$ ($\mathcal{M} < 0$) induces the (inverse) Mpemba effect. This universal thermodynamic framework maps out phase diagram to provide general criteria for the Mpemba effect in complex systems, offering a macroscopic picture that bridges disparate microscopic approaches.

\end{abstract}

\maketitle

\textsl{\textcolor{black}{Introduction.}}\textsl{---} Counter to intuitive expectations, an object at a higher temperature can sometimes cool down faster than a colder one under certain conditions. This thermal phenomenon is known as the Mpemba effect (ME) \citep{mpemba_osborne_1969}. It has fascinated thinkers since antiquity, with early observations recorded by Aristotle, Roger Bacon, Francis Bacon, and Descartes \citep{jeng_2006,burridge_linden_2016}. However, rigorous physical analysis of the ME only began in the late 20th century \citep{mpemba_osborne_1969,kell_1969,firth_1971,deeson_1971,osborne_1979,auerbach_1995,knight_1996,katz_2009,vynnycky_mitchell_2010,balazovic_tomasik_2012,balazovic_tomasik_2015}. The original paradox is intrinsically macroscopic, strictly concerning the observable temperature evolution of bulk matter (e.g., water). Despite decades of research, the underlying macroscopic mechanisms remain debated. Proposed factors, such as evaporation \citep{kell_1969,firth_1971,walker_1977,vynnycky_mitchell_2010}, convection \citep{deeson_1971,maciejewski_1996,vynnycky_maeno_2012,vynnycky_kimura_2015}, supercooling \citep{auerbach_1995,esposito_2008}, nucleation temperature \citep{knight_1996,brownridge_2011} and interfacial phenomena \citep{auerbach_1995,brownridge_2011}, have yielded conflicting results, even raising doubts about the effect's experimental reproducibility in pure water \citep{burridge_linden_2016}.

Faced with both the lack of broadly recognized experimental results to guide phenomenological analysis and the theoretical difficulty of modeling macroscopic systems far from equilibrium, ME research has largely shifted toward the microscopic regime. Mpemba-like anomalous relaxation has been extensively observed across diverse small-scale platforms, including nanotube resonators \citep{greaney_2011}, clathrate hydrates \citep{ahn_2016}, granular fluids \citep{lasanta_2017,torrente_2019}, spin glasses \citep{baity_jesi_2019}, trapped colloidal particles \citep{kumar_2020}, and quantum systems \citep{Chatterjee_2023_PRL,wang2024,Joshi_2024,ares2025quantum}. Concurrently, significant progress in stochastic thermodynamics has led to numerous model-specific studies exploring custom energy landscapes and interactions \citep{esposito_2008,zhang_2014,jin_goddard_2015,greaney_2011,ahn_2016,lasanta_2017,lu_2017,keller_2018,baity_jesi_2019,gijon_2019,nava_fabrizio_2019,torrente_2019,kumar_2020,biswas_maxwell_2020,Vadakkayil_ferromagnet_2021,biswas_brownian_2023}. Various mechanisms have been identified, such as Mpemba indices in Ising models \citep{klich2019mpemba}, non-Markovian memory induced by reservoir back-action \citep{yang2020non}, and the dynamical bypassing of metastable states \citep{zhang2022theoretical}. Notable theoretical milestones include Lu and Raz's framework for Markovian dynamics on complex landscapes \citep{lu_2017}, and Vu and Hayakawa's recent microscopic universal theory based on thermomajorization~\citep{vu_2025}.

Despite these remarkable advances, the prevailing focus on microscopic systems deviates from the key characteristic of the original macroscopic Mpemba phenomenon, namely the crossing of temperature curves~\citep{mpemba_osborne_1969}. Although effective temperature can be defined for specific simple systems~\citep{Chatterjee_2023_PRL, Chatterjee_2024_PRA}, a well-defined temperature is generally absent during far-from-equilibrium relaxation~\citep{TEZA2026}. Consequently, microscopic models predominantly track the ME via statistical geometric approaches~\citep{vu_2025}. Lacking observability, these probabilistic metrics fail to directly capture the crossing of physical cooling trajectories~\citep{mpemba_osborne_1969}. Recognizing this limitation, recent efforts have attempted to revive macroscopic descriptions. For instance, Santos~\citep{santos_2024} successfully reproduced the ME by introducing a phenomenological time delay into Newton's cooling law. However, the thermodynamic origin of such macroscopic memory remains elusive. Ultimately, a critical gap persists: the lack of a universal thermodynamic framework capable of predicting the ME directly through observable temperature trajectories.

To bridge this gap and return the study of anomalous relaxation to its macroscopic origins, we formulate a framework based on linear irreversible thermodynamics (LIT) \citep{prigogine_1947,degroot_1984}, which has successfully established fundamental bounds for heat engines independent of microscopic details~\citep{Broeck_2005,Tu2014,izumida2014work,Yuan2022}, we demonstrate that macroscopic non-Markovian memory arises from the coupling between thermal relaxation and an additional macro-variable. This degree of freedom naturally encompasses reservoir back-action~\citep{ondrechen1981maximum,izumida2014work,richens2018finite,timpanaro2020landauer,Ma2020,Yuan2022,Ma2023} or internal structural evolution of the system. We derive a generalized memory-dependent Newton's cooling law that yields the exact temperature evolution, providing explicit, macroscopically observable criteria for anomalous relaxation.

\textsl{\textcolor{black}{General framework}}\textsl{---} Within the framework of linear irreversible thermodynamics, we establish the fundamental thermodynamic prerequisites for anomalous relaxation from the perspective of total entropy production. Consider a simple system with an instantaneous temperature $T(t)$, immersed in a finite thermal reservoir at $T_{\mathrm{r}}$. Assuming weak coupling, the system-reservoir interaction energy $V$ is negligible compared to the system's internal energy $E$ and the reservoir energy $E_{\mathrm{r}}$. Consequently, the total energy of the isolated composite system is strictly additive and conserved ($E_{\mathrm{tot}} \approx E + E_{\mathrm{r}} = \mathrm{const}$). This conservation constraint dictates that $E_{\mathrm{r}}$ is entirely determined by $E$, leaving which as the sole independent energy parameter. Traditional Newtonian cooling relies exclusively on this single temperature variable, dictating that the cooling time strictly monotonically increases with the initial temperature difference, which fundamentally precludes any anomalous relaxation \citep{lasanta_2017}. To transcend this limitation, we introduce an additional macroscopic variable $\alpha(t)$ for the composite system. Unlike temperature, which captures microscopic random thermal motion, $\alpha$ encapsulates macroscopic configurational or intrinsic changes, thereby serving as a \textit{structural} variable. Crucially, to generate a non-Markovian memory effect, the evolution timescale of $\alpha(t)$ must be comparable to that of the thermal relaxation, preventing it from instantaneously equilibrating with the temperature. With these two independent variables defined, without loss of generality, we write the system's internal energy as $E=E(T, \alpha)$ (seamlessly accommodating cases where $E$ is independent of $\alpha$), and the total entropy of the composite system can be parameterized as $S_{\mathrm{tot}}(E, \alpha)$, yielding a total entropy production rate $\sigma = \mathrm{d}S_{\mathrm{tot}}/\mathrm{d}t$ that reads
\begin{equation}
    \sigma = \frac{\partial S_{\mathrm{tot}}}{\partial E} \frac{\mathrm{d} E}{\mathrm{d} t} + \frac{\partial S_{\mathrm{tot}}}{\partial \alpha} \frac{\mathrm{d} \alpha}{\mathrm{d} t}. \label{eq:entropy_rate}
\end{equation}
Here, the heat flux released by the system is $\mathrm{d}Q/\mathrm{d}t\equiv J_q = -\mathrm{d}E/\mathrm{d}t = -C\mathrm{d}T/\mathrm{d}t - u J_\alpha$, and $J_\alpha = \mathrm{d}\alpha/\mathrm{d}t$, where $C = \partial E/\partial T$ is the heat capacity, and $u = \partial E/\partial \alpha$ represents the internal latent heat associated with structural transitions. The entropy production rate can thus be recast into a positive-definite bilinear form $\sigma = J_q X_q + J_\alpha X_\alpha \ge 0$. Noting that any heat released by the system is entirely absorbed by the reservoir ($\mathrm{d}E_{\mathrm{r}} = -\mathrm{d}E$), the conjugate thermal force $X_q = -\partial S_{\mathrm{tot}}/\partial E=-\left( \partial S_{\mathrm{s}}/\partial E- \partial S_{\mathrm{r}}/\partial E_{\mathrm{r}} \right) =1/ T_{\mathrm{r}}-1/T\approx (T - T_{\mathrm{r}})/T_{\mathrm{r}}^2$. Conjugately, the generalized restorative force is defined as $X_\alpha = \partial S_{\mathrm{tot}}/\partial \alpha$. In the near-equilibrium regime, $X_\alpha$ can be expanded around the equilibrium point $\alpha_{\mathrm{eq}}$ as
\begin{equation}
    X_\alpha(\alpha) \approx X_\alpha(\alpha_{\mathrm{eq}}) + \left( \frac{\partial^2 S_{\mathrm{tot}}}{\partial \alpha^2} \right)_{\alpha_{\mathrm{eq}}} \Delta \alpha = -\frac{k_\alpha}{T_{\mathrm{r}}}\Delta \alpha. \label{eq:X_alpha_expansion}
\end{equation}
where the equilibrium condition $X_\alpha(\alpha_{\mathrm{eq}}) = 0$ have been utilized. Under isothermal reservoir conditions, the total entropy change is related to the total generalized free energy change $\Delta F$ (of the composite system) via $\Delta S_{\mathrm{tot}} = -\Delta F / T_{\mathrm{r}}$. This strict relation defines $k_\alpha = \partial^2 F/\partial \alpha^2$, representing the curvature of the total free energy landscape. 

The macroscopic fluxes thus follow the Onsager phenomenological equations
\begin{align}
    J_q &= L_{qq} X_q + L_{q\alpha} X_\alpha, \label{eq:Onsager1} \\
    J_\alpha &= L_{\alpha q} X_q + L_{\alpha\alpha} X_\alpha. \label{eq:Onsager2}
\end{align}
To characterize the cross-coupling strength between the heat transfer and structural evolution, we introduce the dimensionless coupling strength $q = L_{\alpha q}/\sqrt{L_{qq} L_{\alpha\alpha}}$. Any non-zero cross-coupling ($q \neq 0$) mathematically induces a non-Markovian historical memory (see \textit{Appendix A} of \textbf{End Matter} for details). For simplicity, we consider the \textit{tight-coupling regime} ($|q|=1$), where the determinant of the Onsager matrix vanishes. This imposes a strict constraint $J_\alpha = k J_q$ ($k \equiv \sqrt{L_{\alpha\alpha}/L_{qq}}$), allowing us to rewrite the heat flux decomposition as $J_q = -C\mathrm{d}T/\mathrm{d}t - u k J_q$. Rearranging this defines the temperature evolution governed by the heat flux as
\begin{equation}
    J_q = -\tilde{C} \frac{\mathrm{d}T}{\mathrm{d}t}, \label{eq:Jq_renormalized}
\end{equation}
where thermodynamic stability requires $\tilde{C}\equiv C / (1 + k u) > 0$. Meanwhile,the tight-coupling condition ($L_{q\alpha} = k L_{qq}$) directly simplifies Eq.~(\ref{eq:Onsager1}) to $J_q = L_{qq} (X_q + k X_\alpha)$. To capture structural backaction, we assume the principal transport coefficient is $\alpha$-dependent, expanding to first order as $L_{qq}(\alpha) \approx L_{qq}^{(0)}(1 + \zeta\Delta \alpha)$. Consistently, the structural deviation tracks the cumulative heat via the linear mapping $\Delta \alpha(t) \approx k_0 Q(t)$ \citep{Alpha_Q}, where $\zeta \equiv (\partial L_{qq}/\partial \alpha)/L_{qq}^{(0)}$ and $k_0 \equiv \sqrt{L_{\alpha\alpha}/L_{qq}^{(0)}}$ are treated as constants. Combining these relations gives $J_q = L_{qq}(Q)T_{\mathrm{r}}^{-2} \left[ (T - T_{\mathrm{r}}) - k_0^2 k_\alpha T_{\mathrm{r}} Q(t) \right]$. Equating this with Eq.~(\ref{eq:Jq_renormalized}) and factoring out $\tilde{C}$ leads to the main result of this Letter 
\begin{equation}
    \frac{\mathrm{d}T}{\mathrm{d}t} = - \left[ \gamma_0 + \mathcal{M}Q(t) \right] \big[ (T - T_{\mathrm{r}}) - \mathcal{I} Q(t) \big], \label{eq:master_eq}
\end{equation}
where $\gamma_0 = L_{qq}^{(0)}/(\tilde{C} T_{\mathrm{r}}^2)$ is the bare relaxation rate. Dictated by the cumulative heat transfer, Eq.~(\ref{eq:master_eq}) reveals two independent memory mechanisms: the kinetic coefficient $\mathcal{M} \equiv \gamma_0k_0\zeta$ accelerates cooling (when $\zeta>0$), while the structural inertia $\mathcal{I} = k_0^2 k_\alpha T_{\mathrm{r}}$ quantifies the resistance of the free-energy landscape. Since Eq.~(\ref{eq:master_eq}) naturally recovers the conventional Newton's cooling law in the memoryless limit ($\mathcal{M}=\mathcal{I}=0$), we designate it the \textit{generalized memory-dependent Newton's Cooling Law}. Underscoring its universality, this governing equation is form-invariant for multidimensional $\boldsymbol{\alpha}(t)$ (see Supplemental Material)and seamlessly extends beyond the tight-coupling regime through the systematic rescaling $\mathcal{M} \rightarrow q\mathcal{M}$ (see \textit{Appendix A}).

Notably, the structural inertia $\mathcal{I}$ dictates a residual temperature $\Delta T_{\infty} = \mathcal{I}\tilde{C}\Delta T_0/(1 + \mathcal{I}\tilde{C}) > 0$. This indicates incomplete thermalization and a breakdown of energy equipartition, where fast kinetic modes equilibrate while slow structural modes remain energetically trapped, a phenomenon reminiscent of structural glass transitions~\cite{ballauff_2013,jin_2026}. Because the kinetic response ($\mathcal{M}$) and static landscape ($\mathcal{I}$) are microscopically decoupled, Eq.~(\ref{eq:master_eq}) captures both Mpemba acceleration and metastable trapping. To focus purely on the $\mathcal{M}$-driven Mpemba effect, we assume $\mathcal{I}=0$ hereafter to ensure complete thermalization, deferring the study of incomplete thermalization to future work \citep{IT}.

\begin{figure}[htb!]
    \centering
    \includegraphics[width=\columnwidth]{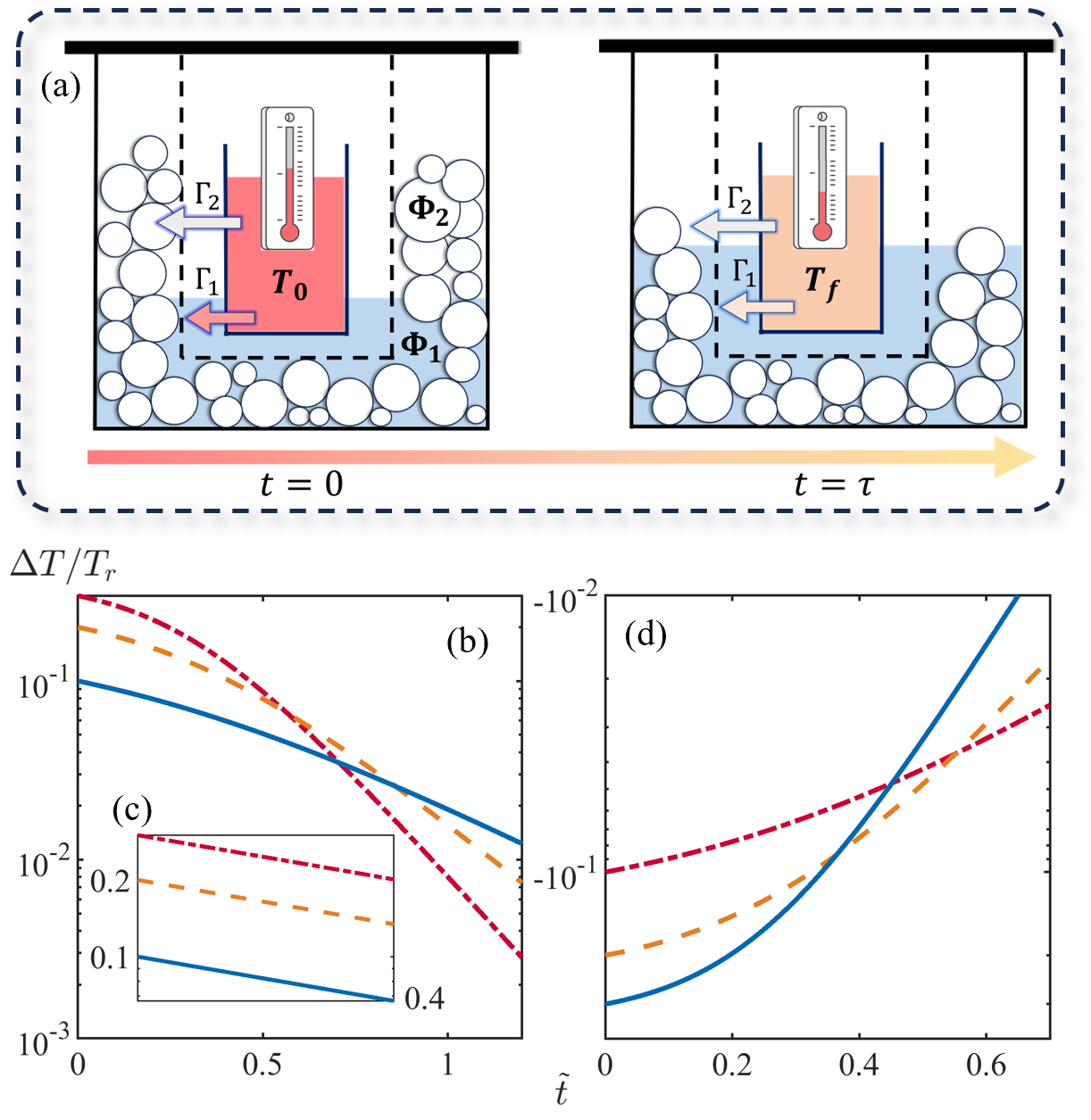}
    \caption{\label{fig:Temperature-different}(a) Schematic of Example I. A system is surrounded by a biphasic reservoir ($T_{\mathrm{r}}$). Heat released by the system drives the phase transition ($\Phi_{2} \to \Phi_{1}$), dynamically enhancing the effective thermal conductance from $\Gamma_2$ (air/ice) to $\Gamma_1$ (water) and satisfying $\mu=\Gamma_1/\Gamma_2>1$. Dimensionless temperature difference $\Delta T/T_r$ as a function of dimensionless time $\tilde{t}=\gamma_0 t$ with $\mu=10$ (b), $\mu=1$ (c) and $\mu=0.06$ (d). The red dash-dotted curve, orange dashed curve, and blue solid curve in (b-c) with $n_{0}=0.1$ are plotted with $\Delta T_0/T_r=0.3$, $\Delta T_0/T_r=0.2$, and $\Delta T_0/T_r=0.1$, respectively. While the red dash-dotted curve, orange dashed curve, and blue solid curve in (d) with $n_{0}=0.9$ are plotted with $\Delta T_0/T_r=-0.1$, $\Delta T_0/T_r=-0.2$, and $\Delta T_0/T_r=-0.3$; other parameters are fixed at $\Gamma_{2}=5$, $C=1$, $CT_r/\Theta=3$, and $\Theta=100$.}
\end{figure}

\textsl{\textcolor{black}{Example I: $\alpha$ as a reservoir variable.}}\textsl{---} It is worth noting that $\alpha$ can originate from different components of the total entropy $S_{\mathrm{tot}}(E, \alpha)$: (i) if $\alpha$ represents a reservoir parameter, heat transfer modifies its macroscopic state; (ii) if $\alpha$ is an internal degree of freedom (e.g., energy landscape complexity), it modulates the system's entropy. We first map this general framework onto a macroscopic scenario where $\alpha(t)$ acts as a reservoir variable (case i). As illustrated in Fig.~\ref{fig:Temperature-different}(a), the reservoir is a biphasic mixture of water ($\Phi_1$) and ice ($\Phi_2$) with fractions $n_1(t)$ and $n_2(t) = 1 - n_1(t)$. For theoretical tractability, we assume an idealized uniform temperature maintained by an ongoing first-order phase transition, neglecting internal thermal gradients. In this limit, heat exchange solely drives inter-phase conversion; so long as $\Phi_2$ persists, the reservoir temperature remains strictly constant without any temperature rise in the water. A system to be cooled is placed inside a meshed container, surrounded by this ice-water reservoir. The mesh separates the ice from the system but allows water to dynamically replace the low-conductance air gap as the ice melts. The contributions of these two phases to the total effective thermal conductance are $\Gamma_{1}$ (water contact) and $\Gamma_{2}$ (air/ice contact) respectively, and thus
\begin{equation}
\Gamma(t) = n_{1}(t)\Gamma_{1} + n_{2}(t)\Gamma_{2}\label{eq:gannae}
\end{equation}

By assuming rapid internal thermalization, where the thermal conductance inside the two phases of the reservoir greatly exceeds the system-reservoir coupling, we can neglect internal temperature gradients and maintain a uniform reservoir temperature $T_{\mathrm{r}}$. With the notation $\Theta$ representing the latent heat required for the reservoir to undergo a complete phase transition, the evolution of the phase population is strictly locked to the heat absorption $\mathrm{d}Q/\mathrm{d}t = -\Gamma(t)(T-T_{\mathrm{r}})$, yielding $\Theta (\mathrm{d}n_{1}/\mathrm{d}t) = -\mathrm{d}Q/\mathrm{d}t$. Integrating this gives
\begin{equation}
n_{1}(t) = n_{0} + \frac{Q(t)}{\Theta}.\label{eq:na}
\end{equation}
Here $n_{0}$ is the initial population of $\Phi_{1}$ and $Q(t)=\tilde{C}\left[T_{0}-T(t)\right]$ is the heat absorbed. Equation~(\ref{eq:na}) perfectly embodies the tight-coupling constraint ($\Delta \alpha = k_0 Q$) of the general framework, with the phase population acting as the macroscopic structural variable $\alpha(t)$ and $k_0 = 1/\Theta$. (Since the variable $\alpha$ belongs strictly to the reservoir here, the system's internal latent heat coupling $u = 0$, thus $\tilde{C} = C$). Combining Eqs. (\ref{eq:gannae}) and (\ref{eq:na}), the time-dependent thermal conductance $\Gamma(t)$ maps exactly to the state-dependent transport $L_{qq}(Q)$. As the ice melts and water dynamically replaces the air gap, the overall heat transfer coefficient is progressively enhanced, naturally resulting $\Gamma_{1}>\Gamma_{2}$. Defining the thermal conductance ratio $\mu\equiv\Gamma_{1}/\Gamma_{2}$, we obtain
\begin{equation}
\Gamma(t)=\Gamma_0+\Theta^{-1}(\mu-1)\Gamma_{2} Q(t) \label{eq:gammat}
\end{equation}
where $\Gamma_0 = [n_{0}(\mu-1)+1]\Gamma_{2}$. By comparing this with the general relaxation rate $\gamma_0 + \mathcal{M}Q(t)$ in Eq.~(\ref{eq:master_eq}), we can strictly identify the typical parameters of this specific model: $\gamma_0 = \Gamma_0/C$, and the memory response coefficient $\mathcal{M} = (\mu-1)\Gamma_2 / (\Theta C)$. In this case, $\mathcal{M}>0$ ($\mu>1$), and the temperature evolution is governed by
\begin{equation}
\frac{\mathrm{d}T}{\mathrm{d}t}=-\left[ \frac{\Gamma_0}{C}+ \frac{(\mu-1)\Gamma_2}{\Theta}(T_{0}-T) \right] \left(T-T_{\mathrm{r}}\right).\label{eq:T-t}
\end{equation}
Solving this equation yields the exact analytical solution for the instantaneous temperature difference $\Delta T(t) = T(t) - T_{\mathrm{r}}$ ($\Delta T_0 \equiv T_0 - T_{\mathrm{r}}$)
\begin{equation}
\frac{\Delta T(t)}{\Delta T_0} = \frac{\gamma_0^{-1}\mathcal{M}C\Delta T_0+1}{\gamma_0^{-1}\mathcal{M}C\Delta T_0+{e^{(\mathcal{M}C\Delta T_0+\gamma_0)t}}}. \label{eq:exact_nonlinear_T}
\end{equation}
Figure~\ref{fig:Temperature-different}(b) plots Eq.~(\ref{eq:exact_nonlinear_T}) against the dimensionless cooling time $\tilde{t}\equiv \gamma_0t$ with $\mu=10$. The crossing of trajectories with different initial temperatures provides direct evidence for the macroscopic ME. For comparison, Fig.~\ref{fig:Temperature-different}(c) illustrates a normal cooling process with $\mu=1$. The parallel nature of these curves on a logarithmic scale rigorously verifies the absence of the multiplicative memory $\mathcal{M}$, thereby precluding the ME.

\begin{figure}
    \centering
\includegraphics[width=\columnwidth]{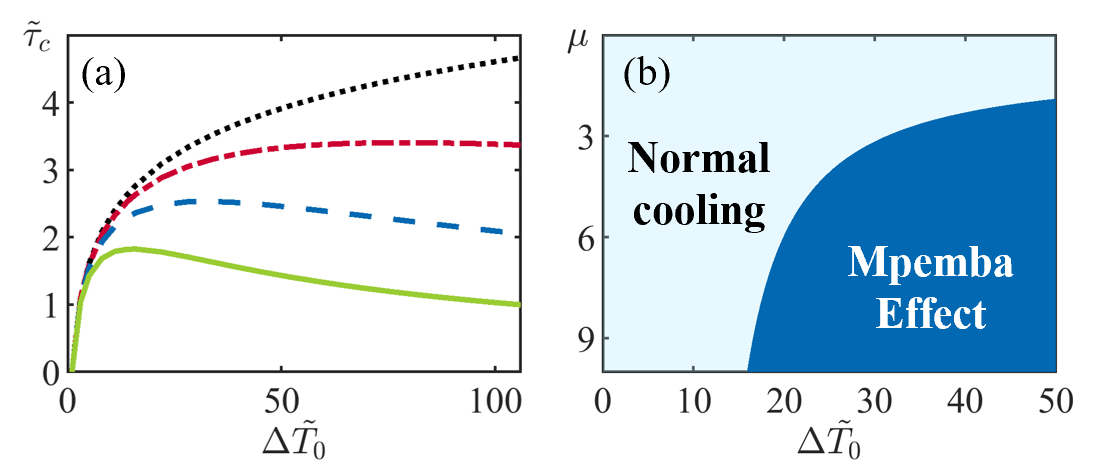}
    \caption{\label{fig:The-cooling-time}The occurrence of the ME. (a) The exact cooling time $\tilde{\tau}_c\equiv\gamma_0\tau_c$ [Eq.~(\ref{eq:exact_tau})] as a function of dimensionless temperature difference $\tilde{\Delta T_0}\equiv\Delta T_0/\Delta T_f$ with $\mu=1,1.5,3,12$ (in order from top to bottom). (b) Phase diagram of the cooling process in the parametric space $(\Delta\tilde{T}_0,\mu)$. The ME region is characterized by $\partial\tau_{c}/\partial T_{0}<0$ while the normal cooling is associated with $\partial\tau_{c}/\partial T_{0}\protect\geq0$.}
\end{figure}

From Eq.~(\ref{eq:exact_nonlinear_T}), the exact cooling time $\tau_c$ to reach a target final temperature $T_f$ [$\Delta T_f = T_f - T_{\mathrm{r}}=\Delta T(\tau_c)$] is analytically given by
\begin{equation}
\tau_c(\Delta T_0) = \frac{\ln \left\{\frac{\Delta T_0}{\Delta T_f} \left[1 + \mathcal{M}C \gamma_0^{-1}(\Delta T_0 - \Delta T_f)\right] \right\}}{\gamma_0 + \mathcal{M}C \Delta T_0}, \label{eq:exact_tau}
\end{equation}
which is plotted in Fig.~\ref{fig:The-cooling-time}(a). Near $T_f$, all curves with different $\mu$ converge and grow linearly as $\tau_c \sim \gamma_0^{-1}(\Delta T_0/\Delta T_f-1)$. Farther from $T_f$, however, the memoryless limit ($\mu=1$, $\mathcal{M}=0$, black dotted curve) uniquely maintains a monotonically increasing trend, scaling as $\tau_c\sim\gamma_0^{-1}\ln(\Delta T_0/\Delta T_f)$. Conversely, any non-zero $\mathcal{M}$ forces $\tau_c$ to decrease in the large-$\Delta T_0$ regime as $\tau_c\sim({\mathcal{M}C \Delta T_0})^{-1}\ln(\Delta T_0/\Delta T_f)$. Thus, the ME is an inevitable consequence of $\mathcal{M} \neq 0$, with larger $\mathcal{M}$ inducing a progressively steeper drop in the cooling time. Dynamically, the direct signature of ME, namely the crossing of cooling curves above $T_f$ [e.g., Fig.~\ref{fig:Temperature-different}(b)], is equivalent to a higher initial temperature yielding a shorter cooling time. This naturally motivates us to define an order parameter $\chi\equiv\partial\tau_{c}/\partial T_{0}$ as a rigorous criterion for the ME. As shown in Fig.~\ref{fig:The-cooling-time}(b), the sign of $\chi$ partitions the 2D parameter space, mapping the boundary between the ME ($\chi<0$) and normal cooling ($\chi>0$) regimes.

Strikingly, our framework also seamlessly unifies the opposite physical extreme where $\mathcal{M} < 0$. During a cooling process ($Q > 0$), a negative $\mathcal{M}$ implies that the dynamically generated structure severely hinders heat transfer ($\gamma_0+\mathcal{M}Q$ decreases). Consequently, the relaxation of a hotter initial state is retarded. A paradigmatic macroscopic manifestation of this cooling slowdown occurs during the quenching of a high-temperature metal in a cold liquid (Leidenfrost effect \citep{vakarelski2012stabilization}): the extreme initial heat flux instantly creates an insulating vapor layer at the interface, plummeting the effective heat conductance and substantially prolonging the cooling time. In our thermodynamic framework, this anomalous cooling retardation of a hotter system is rigorously identified as the \textit{Anti-Mpemba effect} (Anti-ME). Conversely, during a heating process where the system absorbs heat ($Q < 0$), this exact same $\mathcal{M} < 0$ condition dynamically \textit{enhances} the effective heating rate ($-\mathcal{M}Q > 0$). This accelerates the thermalization and strictly triggers the inverse ME, where a colder initial system heats up faster than a warmer one, as illustrated in Fig.~\ref{fig:Temperature-different}(d) with $\mu=0.06$. For a detailed analysis of the inverse ME, please see \textit{Appendix B}.

\textsl{\textcolor{black}{Example II: $\alpha$ as a system variable}}\textsl{---} To explicitly demonstrate the broad universality of our result and its deep connection to previous studies in the realm of stochastic thermodynamics \citep{lu_2017,klich2019mpemba,kumar_2020}, we now map our framework onto a system possessing a complex internal energy landscape (case ii). In microscopic Markovian models, the ME often arises from the probability population dynamically transferring between metastable states and the global minimum. We assume the system possesses fast kinetic/vibrational degrees of freedom that maintain a uniform instantaneous macroscopic temperature $T(t)$. Simultaneously, the system possesses an internal structural degree of freedom coarse-grained into two macrostates: a low-energy ground state ($\varphi_1$) and a higher-energy metastable state ($\varphi_2$) separated by an energy difference $\Omega> 0$. We define the macroscopic structural variable $\alpha(t) \in [0, 1]$ as the population fraction of the metastable state $\varphi_2$. The total internal energy of the system is simply assumed to be $E(T, \alpha) = CT(t) + \alpha(t)\Omega$. This matches our general framework, identifying the internal latent heat as $u =\partial E/\partial \alpha= \Omega$. Due to their distinct macroscopic configurations, it is natural to expect that these two states exhibit different effective thermal conductances to the external reservoir. Let $\Lambda_1$ and $\Lambda_2$ be the conductances of the ground and metastable states, respectively. The total macroscopic heat conductance is the population-weighted average
\begin{equation}
    \Lambda(\alpha) = (1 - \alpha)\Lambda_1 + \alpha \Lambda_2 = \Lambda_1 - \alpha(\Lambda_1 - \Lambda_2). \label{eq:D2}
\end{equation}
During cooling, the system loses energy, driving internal transitions from the high-energy metastable state $\varphi_2$ to the ground state $\varphi_1$ (hence $\Delta \alpha < 0$). Under the tight-coupling condition, this structural evolution is strictly driven by heat dissipation $Q(t)$. We introduce a positive coupling constant $\kappa > 0$, such that the population change is slaved to the cumulative heat released: $\Delta \alpha(t) = - \kappa Q(t)$. Thus, the phenomenological coupling parameter is $k_0 = -\kappa$.

Following our general framework, $\tilde{C} = C / (1 - \kappa\Omega)$. To satisfy thermodynamic stability (temperature must decrease during cooling), we require $1 - \kappa\Omega> 0$, ensuring $\tilde{C} > 0$. Substituting $\alpha(t) = \alpha_0(T_0)- \kappa Q(t)$ into Eq.~\eqref{eq:D2} yields $\Lambda(\alpha) = \Lambda_0(T_0) + \kappa(\Lambda_1 - \Lambda_2)Q(t)$, where $\Lambda_0(T_0)= \Lambda_1 - \alpha_0(T_0)(\Lambda_1 - \Lambda_2)$ is the initial state-dependent conductance. Dividing the primary heat flux $J_q = \Lambda(\alpha)(T - T_{\mathrm{r}})$ by $\tilde{C}$, we arrive directly at our exact non-linear thermal relaxation equation (Eq.~\ref{eq:master_eq} with $\mathcal{I}=0$), perfectly mapping the parameters
\begin{equation}
    \gamma_0(T_0) = \frac{\Lambda_1 - \alpha_0(T_0)(\Lambda_1 - \Lambda_2)}{\tilde{C}}, \quad \mathcal{M}= \frac{\kappa(\Lambda_1 - \Lambda_2)}{\tilde{C}}. 
\end{equation}
Crucially, our macroscopic framework does not rely on any restrictive assumptions regarding the specific ordering of the conductances (i.e., $\Lambda_1 > \Lambda_2$ is not strictly required). Instead, the overall effective cooling exponent evaluated from our exact solution is simply proportional to $\gamma(T_0) = \gamma_0(T_0) + \mathcal{M} \tilde{C} \Delta T_0$. Its derivative with respect to the initial temperature serves as a universal criterion for anomalous relaxation
\begin{equation}
\frac{\partial \gamma }{\partial T_0} = \frac{\Lambda_1 - \Lambda_2}{\tilde{C}} \left[ \kappa \tilde{C} - \frac{\partial \alpha_0}{\partial T_0} \right].   \label{eq:gamma_derivative}
\end{equation}
This demonstrates that the emergence of the Mpemba effect is solely dictated by the sign of the product on the right-hand side. If the two bracketed terms share the same sign (yielding $\partial \gamma/ \partial T_0 > 0$), a hotter initial system dynamically unlocks a faster overall relaxation pathway, triggering the ME. Conversely, if the terms have opposite signs (yielding $\partial \gamma/ \partial T_0 < 0$), the inverse ME can occur.

\textsl{\textcolor{black}{Concluding remarks.}}\textsl{---} In summary, we established a universal macroscopic framework for the Mpemba effect derived strictly from linear irreversible thermodynamics. We demonstrated that anomalous thermal relaxation inherently emerges from the coupling between heat transfer and an additional macroscopic variable. While recent studies attribute the Mpemba effect to specific microscopic details, our framework unifies these diverse observations. As demonstrated in Example II, coarse-graining the complex internal structure successfully captures the system-intrinsic mechanism driving anomalous relaxation. This aligns directly with various microscopic studies identifying internal structural complexity as the fundamental driver of the Mpemba effect, ranging from complex energy landscapes and phase transition behaviors~\citep{baity_jesi_2019,klich2019mpemba,kumar_2020,zhang2022theoretical} to the inherent cross-coupling between heat and particle flux in open quantum systems~\citep{Chatterjee_2023_PRL,wang2024}. Furthermore, our results conceptually bridge these microscopic mechanisms with recent phenomenological macroscopic approaches. While extending Newton's cooling law with an assumed time delay can produce the Mpemba effect phenomenologically~\citep{santos_2024}, our theory unveils the possible physical origin of such historical dependence, rigorously identifying the cumulative heat exchange as the intrinsic memory carrier. Moreover, Example I reveals that anomalous relaxation naturally emerges even when the memory carrier resides solely in the reservoir back-action, consistent with recent numerical observations \citep{yang2020non}. Consequently, our general theory fundamentally unites these previously fragmented microscopic findings.

Since our framework solely requires the memory variable $\alpha$ to modulate the total entropy, $\alpha$ can broadly represent a dynamic parameter characterizing the system-environment interaction. This perspective naturally incorporates classical macroscopic factors historically proposed to explain the Mpemba effect. Specifically, elements such as natural convection or frost formation driven by steep initial temperature gradients \citep{vynnycky_maeno_2012, auerbach_1995} are seamlessly integrated as non-equilibrium boundary memories that modify heat transfer. Crucially, this interpretation accommodates the Leidenfrost effect. Within our framework, the Mpemba and Leidenfrost effects are no longer isolated anomalies but are self-consistently unified as distinct regimes of memory-dependent anomalous relaxation. By analyzing the interplay between the heat flow direction and the sign of the memory response coefficient, we establish a complete taxonomy of macroscopic thermal relaxation in \textit{Appendix B}. This taxonomy rigorously classifies both the ME and inverse ME alongside their corresponding retarded phenomena.

Looking forward, engineering specific system-bath interactions to artificially induce interfacial memory offers novel protocols for accelerating stochastic or quantum thermalization. Furthermore, our generalized memory-dependent heat transfer law opens exciting avenues for exploring and optimally controlling broader non-Markovian thermodynamic processes. Promising theoretical directions include minimizing dissipated work in driven non-equilibrium protocols and optimizing the finite-time performance of thermodynamic cycles that inherently incorporate process memory.

\textsl{\textcolor{black}{Acknowledgments.}}\textsl{---} Y. H. Ma is immensely grateful for the inspiring presentation entitled ''Protocols for the ME in granular fluids'' given by Andres Santos at the STATPHYS28 conference held in Tokyo, Japan. Y. H. Ma also thanks Shiling Liang, Tanji Zhou and Fangming Cui for helpful comments on the original idea of this work. We acknowledge the assistance of Gemini for the help with writing and inspiring discussions. Y.H.M. thanks the National Natural Science Foundation of
China for support under Grant No. 12305037. Z.C.T. thanks the National Natural Science Foundation of China for support under Grant No. 12475032.

\textit{Data availability.---}The data that support the findings of this article are not publicly available upon publication because it is not technically feasible and/or the cost of preparing, depositing, and hosting the data would be prohibitive within the terms of this research project. The data are available from the authors upon reasonable request.
\end{CJK*}
\bibliography{Refs}

\begin{thebibliography}{64}%
\makeatletter
\providecommand \@ifxundefined [1]{%
 \@ifx{#1\undefined}
}%
\providecommand \@ifnum [1]{%
 \ifnum #1\expandafter \@firstoftwo
 \else \expandafter \@secondoftwo
 \fi
}%
\providecommand \@ifx [1]{%
 \ifx #1\expandafter \@firstoftwo
 \else \expandafter \@secondoftwo
 \fi
}%
\providecommand \natexlab [1]{#1}%
\providecommand \enquote  [1]{``#1''}%
\providecommand \bibnamefont  [1]{#1}%
\providecommand \bibfnamefont [1]{#1}%
\providecommand \citenamefont [1]{#1}%
\providecommand \href@noop [0]{\@secondoftwo}%
\providecommand \href [0]{\begingroup \@sanitize@url \@href}%
\providecommand \@href[1]{\@@startlink{#1}\@@href}%
\providecommand \@@href[1]{\endgroup#1\@@endlink}%
\providecommand \@sanitize@url [0]{\catcode `\\12\catcode `\$12\catcode `\&12\catcode `\#12\catcode `\^12\catcode `\_12\catcode `\%12\relax}%
\providecommand \@@startlink[1]{}%
\providecommand \@@endlink[0]{}%
\providecommand \url  [0]{\begingroup\@sanitize@url \@url }%
\providecommand \@url [1]{\endgroup\@href {#1}{\urlprefix }}%
\providecommand \urlprefix  [0]{URL }%
\providecommand \Eprint [0]{\href }%
\providecommand \doibase [0]{https://doi.org/}%
\providecommand \selectlanguage [0]{\@gobble}%
\providecommand \bibinfo  [0]{\@secondoftwo}%
\providecommand \bibfield  [0]{\@secondoftwo}%
\providecommand \translation [1]{[#1]}%
\providecommand \BibitemOpen [0]{}%
\providecommand \bibitemStop [0]{}%
\providecommand \bibitemNoStop [0]{.\EOS\space}%
\providecommand \EOS [0]{\spacefactor3000\relax}%
\providecommand \BibitemShut  [1]{\csname bibitem#1\endcsname}%
\let\auto@bib@innerbib\@empty
\bibitem [{\citenamefont {Mpemba}\ and\ \citenamefont {Osborne}(1969)}]{mpemba_osborne_1969}%
  \BibitemOpen
  \bibfield  {author} {\bibinfo {author} {\bibfnamefont {E.~B.}\ \bibnamefont {Mpemba}}\ and\ \bibinfo {author} {\bibfnamefont {D.~G.}\ \bibnamefont {Osborne}},\ }\bibfield  {title} {\bibinfo {title} {Cool?},\ }\href {https://doi.org/10.1088/0031-9120/4/3/312} {\bibfield  {journal} {\bibinfo  {journal} {Phys. Educ.}\ }\textbf {\bibinfo {volume} {4}},\ \bibinfo {pages} {172} (\bibinfo {year} {1969})}\BibitemShut {NoStop}%
\bibitem [{\citenamefont {Jeng}(2006)}]{jeng_2006}%
  \BibitemOpen
  \bibfield  {author} {\bibinfo {author} {\bibfnamefont {M.}~\bibnamefont {Jeng}},\ }\bibfield  {title} {\bibinfo {title} {The mpemba effect: When can hot water freeze faster than cold?},\ }\href {https://doi.org/10.1119/1.2186331} {\bibfield  {journal} {\bibinfo  {journal} {Am. J. Phys.}\ }\textbf {\bibinfo {volume} {74}},\ \bibinfo {pages} {514} (\bibinfo {year} {2006})}\BibitemShut {NoStop}%
\bibitem [{\citenamefont {Burridge}\ and\ \citenamefont {Linden}(2016)}]{burridge_linden_2016}%
  \BibitemOpen
  \bibfield  {author} {\bibinfo {author} {\bibfnamefont {H.~C.}\ \bibnamefont {Burridge}}\ and\ \bibinfo {author} {\bibfnamefont {P.~F.}\ \bibnamefont {Linden}},\ }\bibfield  {title} {\bibinfo {title} {Questioning the mpemba effect: hot water does not cool more quickly than cold},\ }\href {https://doi.org/10.1038/srep37665} {\bibfield  {journal} {\bibinfo  {journal} {Sci. Rep.}\ }\textbf {\bibinfo {volume} {6}},\ \bibinfo {pages} {37665} (\bibinfo {year} {2016})}\BibitemShut {NoStop}%
\bibitem [{\citenamefont {Kell}(1969)}]{kell_1969}%
  \BibitemOpen
  \bibfield  {author} {\bibinfo {author} {\bibfnamefont {G.~S.}\ \bibnamefont {Kell}},\ }\bibfield  {title} {\bibinfo {title} {The freezing of hot and cold water},\ }\href {https://doi.org/10.1119/1.1975687} {\bibfield  {journal} {\bibinfo  {journal} {Am. J. Phys.}\ }\textbf {\bibinfo {volume} {37}},\ \bibinfo {pages} {564} (\bibinfo {year} {1969})}\BibitemShut {NoStop}%
\bibitem [{\citenamefont {Firth}(1971)}]{firth_1971}%
  \BibitemOpen
  \bibfield  {author} {\bibinfo {author} {\bibfnamefont {I.}~\bibnamefont {Firth}},\ }\bibfield  {title} {\bibinfo {title} {Cooler?},\ }\href {https://doi.org/10.1088/0031-9120/6/1/310} {\bibfield  {journal} {\bibinfo  {journal} {Phys. Educ.}\ }\textbf {\bibinfo {volume} {6}},\ \bibinfo {pages} {32} (\bibinfo {year} {1971})}\BibitemShut {NoStop}%
\bibitem [{\citenamefont {Deeson}(1971)}]{deeson_1971}%
  \BibitemOpen
  \bibfield  {author} {\bibinfo {author} {\bibfnamefont {E.}~\bibnamefont {Deeson}},\ }\bibfield  {title} {\bibinfo {title} {Cooler-lower down},\ }\href {https://doi.org/10.1088/0031-9120/6/1/311} {\bibfield  {journal} {\bibinfo  {journal} {Phys. Educ.}\ }\textbf {\bibinfo {volume} {6}},\ \bibinfo {pages} {42} (\bibinfo {year} {1971})}\BibitemShut {NoStop}%
\bibitem [{\citenamefont {Osborne}(1979)}]{osborne_1979}%
  \BibitemOpen
  \bibfield  {author} {\bibinfo {author} {\bibfnamefont {D.~G.}\ \bibnamefont {Osborne}},\ }\bibfield  {title} {\bibinfo {title} {Mind on ice},\ }\href {https://doi.org/10.1088/0031-9120/14/7/313} {\bibfield  {journal} {\bibinfo  {journal} {Phys. Educ.}\ }\textbf {\bibinfo {volume} {14}},\ \bibinfo {pages} {414} (\bibinfo {year} {1979})}\BibitemShut {NoStop}%
\bibitem [{\citenamefont {Auerbach}(1995)}]{auerbach_1995}%
  \BibitemOpen
  \bibfield  {author} {\bibinfo {author} {\bibfnamefont {D.}~\bibnamefont {Auerbach}},\ }\bibfield  {title} {\bibinfo {title} {Supercooling and the mpemba effect: When hot water freezes quicker than cold},\ }\href {https://doi.org/10.1119/1.18059} {\bibfield  {journal} {\bibinfo  {journal} {Am. J. Phys.}\ }\textbf {\bibinfo {volume} {63}},\ \bibinfo {pages} {882} (\bibinfo {year} {1995})}\BibitemShut {NoStop}%
\bibitem [{\citenamefont {Knight}(1996)}]{knight_1996}%
  \BibitemOpen
  \bibfield  {author} {\bibinfo {author} {\bibfnamefont {C.~A.}\ \bibnamefont {Knight}},\ }\bibfield  {title} {\bibinfo {title} {The mpemba effect: The freezing times of cold and hot water},\ }\href {https://doi.org/10.1119/1.18275} {\bibfield  {journal} {\bibinfo  {journal} {Am. J. Phys.}\ }\textbf {\bibinfo {volume} {64}},\ \bibinfo {pages} {524} (\bibinfo {year} {1996})}\BibitemShut {NoStop}%
\bibitem [{\citenamefont {Katz}(2009)}]{katz_2009}%
  \BibitemOpen
  \bibfield  {author} {\bibinfo {author} {\bibfnamefont {J.~I.}\ \bibnamefont {Katz}},\ }\bibfield  {title} {\bibinfo {title} {When hot water freezes before cold},\ }\href {https://doi.org/10.1119/1.2996187} {\bibfield  {journal} {\bibinfo  {journal} {Am. J. Phys.}\ }\textbf {\bibinfo {volume} {77}},\ \bibinfo {pages} {27} (\bibinfo {year} {2009})}\BibitemShut {NoStop}%
\bibitem [{\citenamefont {Vynnycky}\ and\ \citenamefont {Mitchell}(2010)}]{vynnycky_mitchell_2010}%
  \BibitemOpen
  \bibfield  {author} {\bibinfo {author} {\bibfnamefont {M.}~\bibnamefont {Vynnycky}}\ and\ \bibinfo {author} {\bibfnamefont {S.~L.}\ \bibnamefont {Mitchell}},\ }\bibfield  {title} {\bibinfo {title} {Evaporative cooling and the mpemba effect},\ }\href {https://doi.org/10.1007/s00231-010-0637-z} {\bibfield  {journal} {\bibinfo  {journal} {Heat Mass Transf.}\ }\textbf {\bibinfo {volume} {46}},\ \bibinfo {pages} {881} (\bibinfo {year} {2010})}\BibitemShut {NoStop}%
\bibitem [{\citenamefont {Bal{\'a}{\v z}ovi{\v c}}\ and\ \citenamefont {Tom{\'a}{\v s}ik}(2012)}]{balazovic_tomasik_2012}%
  \BibitemOpen
  \bibfield  {author} {\bibinfo {author} {\bibfnamefont {M.}~\bibnamefont {Bal{\'a}{\v z}ovi{\v c}}}\ and\ \bibinfo {author} {\bibfnamefont {B.}~\bibnamefont {Tom{\'a}{\v s}ik}},\ }\bibfield  {title} {\bibinfo {title} {The mpemba effect, shechtman's quasicrystals and student exploration activities},\ }\href {https://doi.org/10.1088/0031-9120/47/5/568} {\bibfield  {journal} {\bibinfo  {journal} {Phys. Educ.}\ }\textbf {\bibinfo {volume} {47}},\ \bibinfo {pages} {568} (\bibinfo {year} {2012})}\BibitemShut {NoStop}%
\bibitem [{\citenamefont {Bal{\'a}{\v z}ovi{\v c}}\ and\ \citenamefont {Tom{\'a}{\v s}ik}(2015)}]{balazovic_tomasik_2015}%
  \BibitemOpen
  \bibfield  {author} {\bibinfo {author} {\bibfnamefont {M.}~\bibnamefont {Bal{\'a}{\v z}ovi{\v c}}}\ and\ \bibinfo {author} {\bibfnamefont {B.}~\bibnamefont {Tom{\'a}{\v s}ik}},\ }\bibfield  {title} {\bibinfo {title} {Paradox of temperature decreasing without unique explanation},\ }\href {https://doi.org/10.4161/23328940.2014.975576} {\bibfield  {journal} {\bibinfo  {journal} {Temperature}\ }\textbf {\bibinfo {volume} {2}},\ \bibinfo {pages} {61} (\bibinfo {year} {2015})}\BibitemShut {NoStop}%
\bibitem [{\citenamefont {Walker}(1977)}]{walker_1977}%
  \BibitemOpen
  \bibfield  {author} {\bibinfo {author} {\bibfnamefont {J.}~\bibnamefont {Walker}},\ }\bibfield  {title} {\bibinfo {title} {Hot water freezes faster than cold water. why does it do so?},\ }\href {https://doi.org/10.1038/scientificamerican0977-246} {\bibfield  {journal} {\bibinfo  {journal} {Sci. Am.}\ }\textbf {\bibinfo {volume} {237}},\ \bibinfo {pages} {246} (\bibinfo {year} {1977})}\BibitemShut {NoStop}%
\bibitem [{\citenamefont {Maciejewski}(1996)}]{maciejewski_1996}%
  \BibitemOpen
  \bibfield  {author} {\bibinfo {author} {\bibfnamefont {P.~K.}\ \bibnamefont {Maciejewski}},\ }\bibfield  {title} {\bibinfo {title} {Evidence of a convective instability allowing warm water to freeze in less time than cold water},\ }\href {https://doi.org/10.1115/1.2824069} {\bibfield  {journal} {\bibinfo  {journal} {J. Heat Transfer}\ }\textbf {\bibinfo {volume} {118}},\ \bibinfo {pages} {65} (\bibinfo {year} {1996})}\BibitemShut {NoStop}%
\bibitem [{\citenamefont {Vynnycky}\ and\ \citenamefont {Maeno}(2012)}]{vynnycky_maeno_2012}%
  \BibitemOpen
  \bibfield  {author} {\bibinfo {author} {\bibfnamefont {M.}~\bibnamefont {Vynnycky}}\ and\ \bibinfo {author} {\bibfnamefont {N.}~\bibnamefont {Maeno}},\ }\bibfield  {title} {\bibinfo {title} {Axisymmetric natural convection-driven evaporation of hot water and the mpemba effect},\ }\href {https://doi.org/10.1016/j.ijheatmasstransfer.2012.07.060} {\bibfield  {journal} {\bibinfo  {journal} {Int. J. Heat Mass Transf.}\ }\textbf {\bibinfo {volume} {55}},\ \bibinfo {pages} {7297} (\bibinfo {year} {2012})}\BibitemShut {NoStop}%
\bibitem [{\citenamefont {Vynnycky}\ and\ \citenamefont {Kimura}(2015)}]{vynnycky_kimura_2015}%
  \BibitemOpen
  \bibfield  {author} {\bibinfo {author} {\bibfnamefont {M.}~\bibnamefont {Vynnycky}}\ and\ \bibinfo {author} {\bibfnamefont {S.}~\bibnamefont {Kimura}},\ }\bibfield  {title} {\bibinfo {title} {Can natural convection alone explain the mpemba effect?},\ }\href {https://doi.org/10.1016/j.ijheatmasstransfer.2014.09.015} {\bibfield  {journal} {\bibinfo  {journal} {Int. J. Heat Mass Transfer}\ }\textbf {\bibinfo {volume} {80}},\ \bibinfo {pages} {243} (\bibinfo {year} {2015})}\BibitemShut {NoStop}%
\bibitem [{\citenamefont {Esposito}\ \emph {et~al.}(2008)\citenamefont {Esposito}, \citenamefont {De~Risi},\ and\ \citenamefont {Somma}}]{esposito_2008}%
  \BibitemOpen
  \bibfield  {author} {\bibinfo {author} {\bibfnamefont {S.}~\bibnamefont {Esposito}}, \bibinfo {author} {\bibfnamefont {R.}~\bibnamefont {De~Risi}},\ and\ \bibinfo {author} {\bibfnamefont {L.}~\bibnamefont {Somma}},\ }\bibfield  {title} {\bibinfo {title} {Mpemba effect and phase transitions in the adiabatic cooling of water before freezing},\ }\href {https://doi.org/10.1016/j.physa.2007.10.029} {\bibfield  {journal} {\bibinfo  {journal} {Physica A (Amsterdam)}\ }\textbf {\bibinfo {volume} {387}},\ \bibinfo {pages} {757} (\bibinfo {year} {2008})}\BibitemShut {NoStop}%
\bibitem [{\citenamefont {Brownridge}(2011)}]{brownridge_2011}%
  \BibitemOpen
  \bibfield  {author} {\bibinfo {author} {\bibfnamefont {J.~D.}\ \bibnamefont {Brownridge}},\ }\bibfield  {title} {\bibinfo {title} {When does hot water freeze faster then cold water? a search for the mpemba effect},\ }\href {https://doi.org/10.1119/1.3490015} {\bibfield  {journal} {\bibinfo  {journal} {Am. J. Phys.}\ }\textbf {\bibinfo {volume} {79}},\ \bibinfo {pages} {78} (\bibinfo {year} {2011})}\BibitemShut {NoStop}%
\bibitem [{\citenamefont {Greaney}\ \emph {et~al.}(2011)\citenamefont {Greaney}, \citenamefont {Lani}, \citenamefont {Cicero},\ and\ \citenamefont {Grossman}}]{greaney_2011}%
  \BibitemOpen
  \bibfield  {author} {\bibinfo {author} {\bibfnamefont {P.~A.}\ \bibnamefont {Greaney}}, \bibinfo {author} {\bibfnamefont {G.}~\bibnamefont {Lani}}, \bibinfo {author} {\bibfnamefont {G.}~\bibnamefont {Cicero}},\ and\ \bibinfo {author} {\bibfnamefont {J.~C.}\ \bibnamefont {Grossman}},\ }\bibfield  {title} {\bibinfo {title} {Mpemba-like behavior in carbon nanotube resonators},\ }\href {https://doi.org/10.1007/s11661-011-0843-4} {\bibfield  {journal} {\bibinfo  {journal} {Metall. Mater. Trans. A}\ }\textbf {\bibinfo {volume} {42}},\ \bibinfo {pages} {3907} (\bibinfo {year} {2011})}\BibitemShut {NoStop}%
\bibitem [{\citenamefont {Ahn}\ \emph {et~al.}(2016)\citenamefont {Ahn}, \citenamefont {Kang}, \citenamefont {Koh},\ and\ \citenamefont {Lee}}]{ahn_2016}%
  \BibitemOpen
  \bibfield  {author} {\bibinfo {author} {\bibfnamefont {Y.-H.}\ \bibnamefont {Ahn}}, \bibinfo {author} {\bibfnamefont {H.}~\bibnamefont {Kang}}, \bibinfo {author} {\bibfnamefont {D.-Y.}\ \bibnamefont {Koh}},\ and\ \bibinfo {author} {\bibfnamefont {H.}~\bibnamefont {Lee}},\ }\bibfield  {title} {\bibinfo {title} {Experimental verifications of mpemba-like behaviors of clathrate hydrates},\ }\href {https://doi.org/10.1007/s11814-016-0094-1} {\bibfield  {journal} {\bibinfo  {journal} {Korean J. Chem. Eng.}\ }\textbf {\bibinfo {volume} {33}},\ \bibinfo {pages} {1903} (\bibinfo {year} {2016})}\BibitemShut {NoStop}%
\bibitem [{\citenamefont {Lasanta}\ \emph {et~al.}(2017)\citenamefont {Lasanta}, \citenamefont {Reyes}, \citenamefont {Prados},\ and\ \citenamefont {Santos}}]{lasanta_2017}%
  \BibitemOpen
  \bibfield  {author} {\bibinfo {author} {\bibfnamefont {A.}~\bibnamefont {Lasanta}}, \bibinfo {author} {\bibfnamefont {F.~V.}\ \bibnamefont {Reyes}}, \bibinfo {author} {\bibfnamefont {A.}~\bibnamefont {Prados}},\ and\ \bibinfo {author} {\bibfnamefont {A.}~\bibnamefont {Santos}},\ }\bibfield  {title} {\bibinfo {title} {When the hotter cools more quickly: Mpemba effect in granular fluids},\ }\href {https://doi.org/10.1103/PhysRevLett.119.148001} {\bibfield  {journal} {\bibinfo  {journal} {Phys. Rev. Lett.}\ }\textbf {\bibinfo {volume} {119}},\ \bibinfo {pages} {148001} (\bibinfo {year} {2017})}\BibitemShut {NoStop}%
\bibitem [{\citenamefont {Torrente}\ \emph {et~al.}(2019)\citenamefont {Torrente}, \citenamefont {L\'opez-Casta\~no}, \citenamefont {Lasanta}, \citenamefont {Reyes}, \citenamefont {Prados},\ and\ \citenamefont {Santos}}]{torrente_2019}%
  \BibitemOpen
  \bibfield  {author} {\bibinfo {author} {\bibfnamefont {A.}~\bibnamefont {Torrente}}, \bibinfo {author} {\bibfnamefont {M.~A.}\ \bibnamefont {L\'opez-Casta\~no}}, \bibinfo {author} {\bibfnamefont {A.}~\bibnamefont {Lasanta}}, \bibinfo {author} {\bibfnamefont {F.~V.}\ \bibnamefont {Reyes}}, \bibinfo {author} {\bibfnamefont {A.}~\bibnamefont {Prados}},\ and\ \bibinfo {author} {\bibfnamefont {A.}~\bibnamefont {Santos}},\ }\bibfield  {title} {\bibinfo {title} {Large mpemba-like effect in a gas of inelastic rough hard spheres},\ }\href {https://doi.org/10.1103/PhysRevE.99.060901} {\bibfield  {journal} {\bibinfo  {journal} {Phys. Rev. E}\ }\textbf {\bibinfo {volume} {99}},\ \bibinfo {pages} {060901} (\bibinfo {year} {2019})}\BibitemShut {NoStop}%
\bibitem [{\citenamefont {Baity-Jesi}\ \emph {et~al.}(2019)\citenamefont {Baity-Jesi}, \citenamefont {E.}, \citenamefont {A.},\ and\ \citenamefont {D.}}]{baity_jesi_2019}%
  \BibitemOpen
  \bibfield  {author} {\bibinfo {author} {\bibfnamefont {M.}~\bibnamefont {Baity-Jesi}}, \bibinfo {author} {\bibfnamefont {C.}~\bibnamefont {E.}}, \bibinfo {author} {\bibfnamefont {C.}~\bibnamefont {A.}},\ and\ \bibinfo {author} {\bibfnamefont {Y.}~\bibnamefont {D.}},\ }\bibfield  {title} {\bibinfo {title} {The mpemba effect in spin glasses is a persistent memory effect},\ }\href {https://doi.org/10.1073/pnas.1819803116} {\bibfield  {journal} {\bibinfo  {journal} {Proc. Natl Acad. Sci. USA}\ }\textbf {\bibinfo {volume} {116}},\ \bibinfo {pages} {15350} (\bibinfo {year} {2019})}\BibitemShut {NoStop}%
\bibitem [{\citenamefont {Kumar}\ and\ \citenamefont {Bechhoefer}(2020)}]{kumar_2020}%
  \BibitemOpen
  \bibfield  {author} {\bibinfo {author} {\bibfnamefont {A.}~\bibnamefont {Kumar}}\ and\ \bibinfo {author} {\bibfnamefont {J.}~\bibnamefont {Bechhoefer}},\ }\bibfield  {title} {\bibinfo {title} {Exponentially faster cooling in a colloidal system},\ }\href {https://doi.org/10.1038/s41586-020-2560-x} {\bibfield  {journal} {\bibinfo  {journal} {Nature}\ }\textbf {\bibinfo {volume} {584}},\ \bibinfo {pages} {64} (\bibinfo {year} {2020})}\BibitemShut {NoStop}%
\bibitem [{\citenamefont {Chatterjee}\ \emph {et~al.}(2023)\citenamefont {Chatterjee}, \citenamefont {Takada},\ and\ \citenamefont {Hayakawa}}]{Chatterjee_2023_PRL}%
  \BibitemOpen
  \bibfield  {author} {\bibinfo {author} {\bibfnamefont {A.~K.}\ \bibnamefont {Chatterjee}}, \bibinfo {author} {\bibfnamefont {S.}~\bibnamefont {Takada}},\ and\ \bibinfo {author} {\bibfnamefont {H.}~\bibnamefont {Hayakawa}},\ }\bibfield  {title} {\bibinfo {title} {Quantum mpemba effect in a quantum dot with reservoirs},\ }\href {https://doi.org/10.1103/PhysRevLett.131.080402} {\bibfield  {journal} {\bibinfo  {journal} {Phys. Rev. Lett.}\ }\textbf {\bibinfo {volume} {131}},\ \bibinfo {pages} {080402} (\bibinfo {year} {2023})}\BibitemShut {NoStop}%
\bibitem [{\citenamefont {Wang}\ and\ \citenamefont {Wang}(2024)}]{wang2024}%
  \BibitemOpen
  \bibfield  {author} {\bibinfo {author} {\bibfnamefont {X.}~\bibnamefont {Wang}}\ and\ \bibinfo {author} {\bibfnamefont {J.}~\bibnamefont {Wang}},\ }\bibfield  {title} {\bibinfo {title} {Mpemba effects in nonequilibrium open quantum systems},\ }\href {https://doi.org/10.1103/PhysRevResearch.6.033330} {\bibfield  {journal} {\bibinfo  {journal} {Phys. Rev. Res.}\ }\textbf {\bibinfo {volume} {6}},\ \bibinfo {pages} {033330} (\bibinfo {year} {2024})}\BibitemShut {NoStop}%
\bibitem [{\citenamefont {Joshi}\ \emph {et~al.}(2024)\citenamefont {Joshi}, \citenamefont {Franke}, \citenamefont {Rath}, \citenamefont {Ares}, \citenamefont {Murciano}, \citenamefont {Kranzl}, \citenamefont {Blatt}, \citenamefont {Zoller}, \citenamefont {Vermersch}, \citenamefont {Calabrese}, \citenamefont {Roos},\ and\ \citenamefont {Joshi}}]{Joshi_2024}%
  \BibitemOpen
  \bibfield  {author} {\bibinfo {author} {\bibfnamefont {L.~K.}\ \bibnamefont {Joshi}}, \bibinfo {author} {\bibfnamefont {J.}~\bibnamefont {Franke}}, \bibinfo {author} {\bibfnamefont {A.}~\bibnamefont {Rath}}, \bibinfo {author} {\bibfnamefont {F.}~\bibnamefont {Ares}}, \bibinfo {author} {\bibfnamefont {S.}~\bibnamefont {Murciano}}, \bibinfo {author} {\bibfnamefont {F.}~\bibnamefont {Kranzl}}, \bibinfo {author} {\bibfnamefont {R.}~\bibnamefont {Blatt}}, \bibinfo {author} {\bibfnamefont {P.}~\bibnamefont {Zoller}}, \bibinfo {author} {\bibfnamefont {B.}~\bibnamefont {Vermersch}}, \bibinfo {author} {\bibfnamefont {P.}~\bibnamefont {Calabrese}}, \bibinfo {author} {\bibfnamefont {C.~F.}\ \bibnamefont {Roos}},\ and\ \bibinfo {author} {\bibfnamefont {M.~K.}\ \bibnamefont {Joshi}},\ }\bibfield  {title} {\bibinfo {title} {Observing the quantum mpemba effect in quantum simulations},\ }\href {https://doi.org/10.1103/PhysRevLett.133.010402} {\bibfield  {journal} {\bibinfo  {journal} {Phys. Rev. Lett.}\ }\textbf {\bibinfo
  {volume} {133}},\ \bibinfo {pages} {010402} (\bibinfo {year} {2024})}\BibitemShut {NoStop}%
\bibitem [{\citenamefont {Ares}\ \emph {et~al.}(2025)\citenamefont {Ares}, \citenamefont {Calabrese},\ and\ \citenamefont {Murciano}}]{ares2025quantum}%
  \BibitemOpen
  \bibfield  {author} {\bibinfo {author} {\bibfnamefont {F.}~\bibnamefont {Ares}}, \bibinfo {author} {\bibfnamefont {P.}~\bibnamefont {Calabrese}},\ and\ \bibinfo {author} {\bibfnamefont {S.}~\bibnamefont {Murciano}},\ }\bibfield  {title} {\bibinfo {title} {The quantum mpemba effects},\ }\href {https://doi.org/10.1038/s42254-025-00838-0} {\bibfield  {journal} {\bibinfo  {journal} {Nature Reviews Physics}\ }\textbf {\bibinfo {volume} {7}},\ \bibinfo {pages} {451} (\bibinfo {year} {2025})}\BibitemShut {NoStop}%
\bibitem [{\citenamefont {Zhang}\ \emph {et~al.}(2014)\citenamefont {Zhang}, \citenamefont {Huang}, \citenamefont {Ma}, \citenamefont {Zhou}, \citenamefont {Zhou}, \citenamefont {Zheng}, \citenamefont {Jiang},\ and\ \citenamefont {Sun}}]{zhang_2014}%
  \BibitemOpen
  \bibfield  {author} {\bibinfo {author} {\bibfnamefont {X.}~\bibnamefont {Zhang}}, \bibinfo {author} {\bibfnamefont {Y.}~\bibnamefont {Huang}}, \bibinfo {author} {\bibfnamefont {Z.}~\bibnamefont {Ma}}, \bibinfo {author} {\bibfnamefont {Y.}~\bibnamefont {Zhou}}, \bibinfo {author} {\bibfnamefont {J.}~\bibnamefont {Zhou}}, \bibinfo {author} {\bibfnamefont {W.}~\bibnamefont {Zheng}}, \bibinfo {author} {\bibfnamefont {Q.}~\bibnamefont {Jiang}},\ and\ \bibinfo {author} {\bibfnamefont {C.~Q.}\ \bibnamefont {Sun}},\ }\bibfield  {title} {\bibinfo {title} {Hydrogen-bond memory and water-skin supersolidity resolving the mpemba paradox},\ }\href {https://doi.org/10.1039/C4CP03669G} {\bibfield  {journal} {\bibinfo  {journal} {Phys. Chem. Chem. Phys.}\ }\textbf {\bibinfo {volume} {16}},\ \bibinfo {pages} {22995} (\bibinfo {year} {2014})}\BibitemShut {NoStop}%
\bibitem [{\citenamefont {Jin}\ and\ \citenamefont {Goddard}(2015)}]{jin_goddard_2015}%
  \BibitemOpen
  \bibfield  {author} {\bibinfo {author} {\bibfnamefont {J.}~\bibnamefont {Jin}}\ and\ \bibinfo {author} {\bibfnamefont {W.~A.}\ \bibnamefont {Goddard}},\ }\bibfield  {title} {\bibinfo {title} {Mechanisms underlying the mpemba effect in water from molecular dynamics simulations},\ }\href {https://doi.org/10.1021/jp511752n} {\bibfield  {journal} {\bibinfo  {journal} {J. Phys. Chem. C}\ }\textbf {\bibinfo {volume} {119}},\ \bibinfo {pages} {2622} (\bibinfo {year} {2015})}\BibitemShut {NoStop}%
\bibitem [{\citenamefont {Lu}\ and\ \citenamefont {Raz}(2017)}]{lu_2017}%
  \BibitemOpen
  \bibfield  {author} {\bibinfo {author} {\bibfnamefont {Z.}~\bibnamefont {Lu}}\ and\ \bibinfo {author} {\bibfnamefont {O.}~\bibnamefont {Raz}},\ }\bibfield  {title} {\bibinfo {title} {Nonequilibrium thermodynamics of the markovian mpemba effect and its inverse},\ }\href {https://doi.org/10.1073/pnas.1701264114} {\bibfield  {journal} {\bibinfo  {journal} {Proc. Natl Acad. Sci. USA}\ }\textbf {\bibinfo {volume} {114}},\ \bibinfo {pages} {5083} (\bibinfo {year} {2017})}\BibitemShut {NoStop}%
\bibitem [{\citenamefont {Keller}\ \emph {et~al.}(2018)\citenamefont {Keller}, \citenamefont {Torggler}, \citenamefont {J{\"a}ger}, \citenamefont {Sch{\"u}tz}, \citenamefont {Ritsch},\ and\ \citenamefont {Morigi}}]{keller_2018}%
  \BibitemOpen
  \bibfield  {author} {\bibinfo {author} {\bibfnamefont {T.}~\bibnamefont {Keller}}, \bibinfo {author} {\bibfnamefont {V.}~\bibnamefont {Torggler}}, \bibinfo {author} {\bibfnamefont {S.~B.}\ \bibnamefont {J{\"a}ger}}, \bibinfo {author} {\bibfnamefont {S.}~\bibnamefont {Sch{\"u}tz}}, \bibinfo {author} {\bibfnamefont {H.}~\bibnamefont {Ritsch}},\ and\ \bibinfo {author} {\bibfnamefont {G.}~\bibnamefont {Morigi}},\ }\bibfield  {title} {\bibinfo {title} {Quenches across the self-organization transition in multimode cavities},\ }\href {https://doi.org/10.1088/1367-2630/aaa161} {\bibfield  {journal} {\bibinfo  {journal} {New J. Phys.}\ }\textbf {\bibinfo {volume} {20}},\ \bibinfo {pages} {025004} (\bibinfo {year} {2018})}\BibitemShut {NoStop}%
\bibitem [{\citenamefont {Gij\'on}\ \emph {et~al.}(2019)\citenamefont {Gij\'on}, \citenamefont {Lasanta},\ and\ \citenamefont {Hern\'andez}}]{gijon_2019}%
  \BibitemOpen
  \bibfield  {author} {\bibinfo {author} {\bibfnamefont {A.}~\bibnamefont {Gij\'on}}, \bibinfo {author} {\bibfnamefont {A.}~\bibnamefont {Lasanta}},\ and\ \bibinfo {author} {\bibfnamefont {E.~R.}\ \bibnamefont {Hern\'andez}},\ }\bibfield  {title} {\bibinfo {title} {Paths towards equilibrium in molecular systems: The case of water},\ }\href {https://doi.org/10.1103/PhysRevE.100.032103} {\bibfield  {journal} {\bibinfo  {journal} {Phys. Rev. E}\ }\textbf {\bibinfo {volume} {100}},\ \bibinfo {pages} {032103} (\bibinfo {year} {2019})}\BibitemShut {NoStop}%
\bibitem [{\citenamefont {Nava}\ and\ \citenamefont {Fabrizio}(2019)}]{nava_fabrizio_2019}%
  \BibitemOpen
  \bibfield  {author} {\bibinfo {author} {\bibfnamefont {A.}~\bibnamefont {Nava}}\ and\ \bibinfo {author} {\bibfnamefont {M.}~\bibnamefont {Fabrizio}},\ }\bibfield  {title} {\bibinfo {title} {Lindblad dissipative dynamics in the presence of phase coexistence},\ }\href {https://doi.org/10.1103/PhysRevB.100.125102} {\bibfield  {journal} {\bibinfo  {journal} {Phys. Rev. B}\ }\textbf {\bibinfo {volume} {100}},\ \bibinfo {pages} {125102} (\bibinfo {year} {2019})}\BibitemShut {NoStop}%
\bibitem [{\citenamefont {Biswas}\ \emph {et~al.}(2020)\citenamefont {Biswas}, \citenamefont {Prasad}, \citenamefont {Raz},\ and\ \citenamefont {Rajesh}}]{biswas_maxwell_2020}%
  \BibitemOpen
  \bibfield  {author} {\bibinfo {author} {\bibfnamefont {A.}~\bibnamefont {Biswas}}, \bibinfo {author} {\bibfnamefont {V.~V.}\ \bibnamefont {Prasad}}, \bibinfo {author} {\bibfnamefont {O.}~\bibnamefont {Raz}},\ and\ \bibinfo {author} {\bibfnamefont {R.}~\bibnamefont {Rajesh}},\ }\bibfield  {title} {\bibinfo {title} {Mpemba effect in driven granular maxwell gases},\ }\href {https://doi.org/10.1103/PhysRevE.102.012906} {\bibfield  {journal} {\bibinfo  {journal} {Phys. Rev. E}\ }\textbf {\bibinfo {volume} {102}},\ \bibinfo {pages} {012906} (\bibinfo {year} {2020})}\BibitemShut {NoStop}%
\bibitem [{\citenamefont {Vadakkayil}\ and\ \citenamefont {Das}(2021)}]{Vadakkayil_ferromagnet_2021}%
  \BibitemOpen
  \bibfield  {author} {\bibinfo {author} {\bibfnamefont {N.}~\bibnamefont {Vadakkayil}}\ and\ \bibinfo {author} {\bibfnamefont {S.~K.}\ \bibnamefont {Das}},\ }\bibfield  {title} {\bibinfo {title} {Should a hotter paramagnet transform quicker to a ferromagnet? monte carlo simulation results for ising model},\ }\href {https://doi.org/10.1039/D1CP00879J} {\bibfield  {journal} {\bibinfo  {journal} {Phys. Chem. Chem. Phys.}\ }\textbf {\bibinfo {volume} {23}},\ \bibinfo {pages} {11186} (\bibinfo {year} {2021})}\BibitemShut {NoStop}%
\bibitem [{\citenamefont {Biswas}\ and\ \citenamefont {Rajesh}(2023)}]{biswas_brownian_2023}%
  \BibitemOpen
  \bibfield  {author} {\bibinfo {author} {\bibfnamefont {A.}~\bibnamefont {Biswas}}\ and\ \bibinfo {author} {\bibfnamefont {R.}~\bibnamefont {Rajesh}},\ }\bibfield  {title} {\bibinfo {title} {Mpemba effect for a brownian particle trapped in a single well potential},\ }\href {https://doi.org/10.1103/PhysRevE.108.024131} {\bibfield  {journal} {\bibinfo  {journal} {Phys. Rev. E}\ }\textbf {\bibinfo {volume} {108}},\ \bibinfo {pages} {024131} (\bibinfo {year} {2023})}\BibitemShut {NoStop}%
\bibitem [{\citenamefont {Klich}\ \emph {et~al.}(2019)\citenamefont {Klich}, \citenamefont {Raz}, \citenamefont {Hirschberg},\ and\ \citenamefont {Vucelja}}]{klich2019mpemba}%
  \BibitemOpen
  \bibfield  {author} {\bibinfo {author} {\bibfnamefont {I.}~\bibnamefont {Klich}}, \bibinfo {author} {\bibfnamefont {O.}~\bibnamefont {Raz}}, \bibinfo {author} {\bibfnamefont {O.}~\bibnamefont {Hirschberg}},\ and\ \bibinfo {author} {\bibfnamefont {M.}~\bibnamefont {Vucelja}},\ }\bibfield  {title} {\bibinfo {title} {Mpemba index and anomalous relaxation},\ }\href {https://doi.org/10.1103/PhysRevX.9.021060} {\bibfield  {journal} {\bibinfo  {journal} {Phys. Rev. X}\ }\textbf {\bibinfo {volume} {9}},\ \bibinfo {pages} {021060} (\bibinfo {year} {2019})}\BibitemShut {NoStop}%
\bibitem [{\citenamefont {Yang}\ and\ \citenamefont {Hou}(2020)}]{yang2020non}%
  \BibitemOpen
  \bibfield  {author} {\bibinfo {author} {\bibfnamefont {Z.-Y.}\ \bibnamefont {Yang}}\ and\ \bibinfo {author} {\bibfnamefont {J.-X.}\ \bibnamefont {Hou}},\ }\bibfield  {title} {\bibinfo {title} {Non-markovian mpemba effect in mean-field systems},\ }\href {https://doi.org/10.1103/PhysRevE.101.052106} {\bibfield  {journal} {\bibinfo  {journal} {Phys. Rev. E}\ }\textbf {\bibinfo {volume} {101}},\ \bibinfo {pages} {052106} (\bibinfo {year} {2020})}\BibitemShut {NoStop}%
\bibitem [{\citenamefont {Zhang}\ and\ \citenamefont {Hou}(2022)}]{zhang2022theoretical}%
  \BibitemOpen
  \bibfield  {author} {\bibinfo {author} {\bibfnamefont {S.}~\bibnamefont {Zhang}}\ and\ \bibinfo {author} {\bibfnamefont {J.-X.}\ \bibnamefont {Hou}},\ }\bibfield  {title} {\bibinfo {title} {Theoretical model for the mpemba effect through the canonical first-order phase transition},\ }\href {https://doi.org/10.1103/PhysRevE.106.034131} {\bibfield  {journal} {\bibinfo  {journal} {Phys. Rev. E}\ }\textbf {\bibinfo {volume} {106}},\ \bibinfo {pages} {034131} (\bibinfo {year} {2022})}\BibitemShut {NoStop}%
\bibitem [{\citenamefont {Van~Vu}\ and\ \citenamefont {Hayakawa}(2025)}]{vu_2025}%
  \BibitemOpen
  \bibfield  {author} {\bibinfo {author} {\bibfnamefont {T.}~\bibnamefont {Van~Vu}}\ and\ \bibinfo {author} {\bibfnamefont {H.}~\bibnamefont {Hayakawa}},\ }\bibfield  {title} {\bibinfo {title} {Thermomajorization mpemba effect},\ }\href {https://doi.org/10.1103/PhysRevLett.134.107101} {\bibfield  {journal} {\bibinfo  {journal} {Phys. Rev. Lett.}\ }\textbf {\bibinfo {volume} {134}},\ \bibinfo {pages} {107101} (\bibinfo {year} {2025})}\BibitemShut {NoStop}%
\bibitem [{\citenamefont {Chatterjee}\ \emph {et~al.}(2024)\citenamefont {Chatterjee}, \citenamefont {Takada},\ and\ \citenamefont {Hayakawa}}]{Chatterjee_2024_PRA}%
  \BibitemOpen
  \bibfield  {author} {\bibinfo {author} {\bibfnamefont {A.~K.}\ \bibnamefont {Chatterjee}}, \bibinfo {author} {\bibfnamefont {S.}~\bibnamefont {Takada}},\ and\ \bibinfo {author} {\bibfnamefont {H.}~\bibnamefont {Hayakawa}},\ }\bibfield  {title} {\bibinfo {title} {Multiple quantum mpemba effect: Exceptional points and oscillations},\ }\href {https://doi.org/10.1103/PhysRevA.110.022213} {\bibfield  {journal} {\bibinfo  {journal} {Phys. Rev. A}\ }\textbf {\bibinfo {volume} {110}},\ \bibinfo {pages} {022213} (\bibinfo {year} {2024})}\BibitemShut {NoStop}%
\bibitem [{\citenamefont {Teza}\ \emph {et~al.}(2026)\citenamefont {Teza}, \citenamefont {Bechhoefer}, \citenamefont {Lasanta}, \citenamefont {Raz},\ and\ \citenamefont {Vucelja}}]{TEZA2026}%
  \BibitemOpen
  \bibfield  {author} {\bibinfo {author} {\bibfnamefont {G.}~\bibnamefont {Teza}}, \bibinfo {author} {\bibfnamefont {J.}~\bibnamefont {Bechhoefer}}, \bibinfo {author} {\bibfnamefont {A.}~\bibnamefont {Lasanta}}, \bibinfo {author} {\bibfnamefont {O.}~\bibnamefont {Raz}},\ and\ \bibinfo {author} {\bibfnamefont {M.}~\bibnamefont {Vucelja}},\ }\bibfield  {title} {\bibinfo {title} {Speedups in nonequilibrium thermal relaxation: Mpemba and related effects},\ }\href {https://doi.org/https://doi.org/10.1016/j.physrep.2025.10.009} {\bibfield  {journal} {\bibinfo  {journal} {Physics Reports}\ }\textbf {\bibinfo {volume} {1164}},\ \bibinfo {pages} {1} (\bibinfo {year} {2026})}\BibitemShut {NoStop}%
\bibitem [{\citenamefont {Santos}(2024)}]{santos_2024}%
  \BibitemOpen
  \bibfield  {author} {\bibinfo {author} {\bibfnamefont {A.}~\bibnamefont {Santos}},\ }\bibfield  {title} {\bibinfo {title} {Mpemba meets newton: Exploring the mpemba and kovacs effects in the time-delayed cooling law},\ }\href {https://doi.org/10.1103/PhysRevE.109.044149} {\bibfield  {journal} {\bibinfo  {journal} {Phys. Rev. E}\ }\textbf {\bibinfo {volume} {109}},\ \bibinfo {pages} {044149} (\bibinfo {year} {2024})}\BibitemShut {NoStop}%
\bibitem [{\citenamefont {Prigogine}(1947)}]{prigogine_1947}%
  \BibitemOpen
  \bibfield  {author} {\bibinfo {author} {\bibfnamefont {I.}~\bibnamefont {Prigogine}},\ }\href@noop {} {\emph {\bibinfo {title} {{\'{E}tude thermodynamique des ph\'{e}nom\`{e}nes irr\'{e}versibles}}}}\ (\bibinfo  {publisher} {Desoer},\ \bibinfo {address} {Li\`{e}ge},\ \bibinfo {year} {1947})\BibitemShut {NoStop}%
\bibitem [{\citenamefont {de~Groot}\ and\ \citenamefont {Mazur}(1984)}]{degroot_1984}%
  \BibitemOpen
  \bibfield  {author} {\bibinfo {author} {\bibfnamefont {S.~R.}\ \bibnamefont {de~Groot}}\ and\ \bibinfo {author} {\bibfnamefont {P.}~\bibnamefont {Mazur}},\ }\href@noop {} {\emph {\bibinfo {title} {{Non-Equilibrium Thermodynamics}}}}\ (\bibinfo  {publisher} {Dover Publications},\ \bibinfo {address} {New York},\ \bibinfo {year} {1984})\BibitemShut {NoStop}%
\bibitem [{\citenamefont {Van~den Broeck}(2005)}]{Broeck_2005}%
  \BibitemOpen
  \bibfield  {author} {\bibinfo {author} {\bibfnamefont {C.}~\bibnamefont {Van~den Broeck}},\ }\bibfield  {title} {\bibinfo {title} {Thermodynamic efficiency at maximum power},\ }\href {https://doi.org/10.1103/PhysRevLett.95.190602} {\bibfield  {journal} {\bibinfo  {journal} {Phys. Rev. Lett.}\ }\textbf {\bibinfo {volume} {95}},\ \bibinfo {pages} {190602} (\bibinfo {year} {2005})}\BibitemShut {NoStop}%
\bibitem [{\citenamefont {Sheng}\ and\ \citenamefont {Tu}(2014)}]{Tu2014}%
  \BibitemOpen
  \bibfield  {author} {\bibinfo {author} {\bibfnamefont {S.}~\bibnamefont {Sheng}}\ and\ \bibinfo {author} {\bibfnamefont {Z.~C.}\ \bibnamefont {Tu}},\ }\bibfield  {title} {\bibinfo {title} {Weighted reciprocal of temperature, weighted thermal flux, and their applications in finite-time thermodynamics},\ }\href {https://doi.org/10.1103/PhysRevE.89.012129} {\bibfield  {journal} {\bibinfo  {journal} {Phys. Rev. E}\ }\textbf {\bibinfo {volume} {89}},\ \bibinfo {pages} {012129} (\bibinfo {year} {2014})}\BibitemShut {NoStop}%
\bibitem [{\citenamefont {Izumida}\ and\ \citenamefont {Okuda}(2014)}]{izumida2014work}%
  \BibitemOpen
  \bibfield  {author} {\bibinfo {author} {\bibfnamefont {Y.}~\bibnamefont {Izumida}}\ and\ \bibinfo {author} {\bibfnamefont {K.}~\bibnamefont {Okuda}},\ }\bibfield  {title} {\bibinfo {title} {Work output and efficiency at maximum power of linear irreversible heat engines operating with a finite-sized heat source},\ }\href {https://doi.org/10.1103/PhysRevLett.112.180603} {\bibfield  {journal} {\bibinfo  {journal} {Phys. Rev. Lett.}\ }\textbf {\bibinfo {volume} {112}},\ \bibinfo {pages} {180603} (\bibinfo {year} {2014})}\BibitemShut {NoStop}%
\bibitem [{\citenamefont {Yuan}\ \emph {et~al.}(2022)\citenamefont {Yuan}, \citenamefont {Ma},\ and\ \citenamefont {Sun}}]{Yuan2022}%
  \BibitemOpen
  \bibfield  {author} {\bibinfo {author} {\bibfnamefont {H.}~\bibnamefont {Yuan}}, \bibinfo {author} {\bibfnamefont {Y.~H.}\ \bibnamefont {Ma}},\ and\ \bibinfo {author} {\bibfnamefont {C.~P.}\ \bibnamefont {Sun}},\ }\bibfield  {title} {\bibinfo {title} {Optimizing thermodynamic cycles with two finite-sized reservoirs},\ }\href {https://doi.org/10.1103/PhysRevE.105.L022101} {\bibfield  {journal} {\bibinfo  {journal} {Phys. Rev. E}\ }\textbf {\bibinfo {volume} {105}},\ \bibinfo {pages} {L022101} (\bibinfo {year} {2022})}\BibitemShut {NoStop}%
\bibitem [{\citenamefont {Ondrechen}\ \emph {et~al.}(1981)\citenamefont {Ondrechen}, \citenamefont {Andresen}, \citenamefont {Mozurkewich},\ and\ \citenamefont {Berry}}]{ondrechen1981maximum}%
  \BibitemOpen
  \bibfield  {author} {\bibinfo {author} {\bibfnamefont {M.~J.}\ \bibnamefont {Ondrechen}}, \bibinfo {author} {\bibfnamefont {B.}~\bibnamefont {Andresen}}, \bibinfo {author} {\bibfnamefont {M.}~\bibnamefont {Mozurkewich}},\ and\ \bibinfo {author} {\bibfnamefont {R.~S.}\ \bibnamefont {Berry}},\ }\bibfield  {title} {\bibinfo {title} {Maximum work from a finite reservoir by sequential carnot cycles},\ }\href {https://doi.org/10.1119/1.12426} {\bibfield  {journal} {\bibinfo  {journal} {Am. J. Phys.}\ }\textbf {\bibinfo {volume} {49}},\ \bibinfo {pages} {681} (\bibinfo {year} {1981})}\BibitemShut {NoStop}%
\bibitem [{\citenamefont {Richens}\ \emph {et~al.}(2018)\citenamefont {Richens}, \citenamefont {Alhambra},\ and\ \citenamefont {Masanes}}]{richens2018finite}%
  \BibitemOpen
  \bibfield  {author} {\bibinfo {author} {\bibfnamefont {J.~G.}\ \bibnamefont {Richens}}, \bibinfo {author} {\bibfnamefont {A.~M.}\ \bibnamefont {Alhambra}},\ and\ \bibinfo {author} {\bibfnamefont {L.}~\bibnamefont {Masanes}},\ }\bibfield  {title} {\bibinfo {title} {Finite-bath corrections to the second law of thermodynamics},\ }\href {https://doi.org/10.1103/PhysRevE.97.062132} {\bibfield  {journal} {\bibinfo  {journal} {Phys. Rev. E}\ }\textbf {\bibinfo {volume} {97}},\ \bibinfo {pages} {062132} (\bibinfo {year} {2018})}\BibitemShut {NoStop}%
\bibitem [{\citenamefont {Timpanaro}\ \emph {et~al.}(2020)\citenamefont {Timpanaro}, \citenamefont {Santos},\ and\ \citenamefont {Landi}}]{timpanaro2020landauer}%
  \BibitemOpen
  \bibfield  {author} {\bibinfo {author} {\bibfnamefont {A.~M.}\ \bibnamefont {Timpanaro}}, \bibinfo {author} {\bibfnamefont {J.~P.}\ \bibnamefont {Santos}},\ and\ \bibinfo {author} {\bibfnamefont {G.~T.}\ \bibnamefont {Landi}},\ }\bibfield  {title} {\bibinfo {title} {Landauer's principle at zero temperature},\ }\href {https://doi.org/10.1103/PhysRevLett.124.240601} {\bibfield  {journal} {\bibinfo  {journal} {Phys. Rev. Lett.}\ }\textbf {\bibinfo {volume} {124}},\ \bibinfo {pages} {240601} (\bibinfo {year} {2020})}\BibitemShut {NoStop}%
\bibitem [{\citenamefont {Ma}(2020)}]{Ma2020}%
  \BibitemOpen
  \bibfield  {author} {\bibinfo {author} {\bibfnamefont {Y.~H.}\ \bibnamefont {Ma}},\ }\bibfield  {title} {\bibinfo {title} {Effect of finite-size heat source's heat capacity on the efficiency of heat engine},\ }\href {https://doi.org/10.3390/e22091002} {\bibfield  {journal} {\bibinfo  {journal} {Entropy}\ }\textbf {\bibinfo {volume} {22}},\ \bibinfo {pages} {1002} (\bibinfo {year} {2020})}\BibitemShut {NoStop}%
\bibitem [{\citenamefont {Ma}(2023)}]{Ma2023}%
  \BibitemOpen
  \bibfield  {author} {\bibinfo {author} {\bibfnamefont {Y.~H.}\ \bibnamefont {Ma}},\ }\bibfield  {title} {\bibinfo {title} {Simple realization of the polytropic process with a finite-sized reservoir},\ }\href {https://doi.org/10.1119/5.0104382} {\bibfield  {journal} {\bibinfo  {journal} {Am. J. Phys.}\ }\textbf {\bibinfo {volume} {91}},\ \bibinfo {pages} {555} (\bibinfo {year} {2023})}\BibitemShut {NoStop}%
\bibitem [{Alpha_Q()}]{Alpha_Q}%
  \BibitemOpen
  \bibinfo {note} {Because $k$ is intrinsically tied to $L_{qq}(\alpha)$, integrating the exact kinetic relation $\mathrm{d}\alpha = k(\alpha) \mathrm{d}Q=\sqrt{L_{\alpha\alpha} / L_{qq}(\alpha)}\mathrm{d}Q$ yields $\Delta \alpha(t) = \zeta^{-1}\{[1 + 3k_0\zeta Q(t)/2]^{2/3} - 1\}$, expanding which for small structural deviations, $\zeta \Delta \alpha \ll 1$, yields the linear relation $\Delta \alpha(t) \approx k_0 Q(t)$}\BibitemShut {NoStop}%
\bibitem [{\citenamefont {Ballauff}\ \emph {et~al.}(2013)\citenamefont {Ballauff}, \citenamefont {Brader}, \citenamefont {Egelhaaf}, \citenamefont {Fuchs}, \citenamefont {Horbach}, \citenamefont {Koumakis}, \citenamefont {Kr\"uger}, \citenamefont {Laurati}, \citenamefont {Mutch}, \citenamefont {Petekidis}, \citenamefont {Siebenb\"urger}, \citenamefont {Voigtmann},\ and\ \citenamefont {Zausch}}]{ballauff_2013}%
  \BibitemOpen
  \bibfield  {author} {\bibinfo {author} {\bibfnamefont {M.}~\bibnamefont {Ballauff}}, \bibinfo {author} {\bibfnamefont {J.~M.}\ \bibnamefont {Brader}}, \bibinfo {author} {\bibfnamefont {S.~U.}\ \bibnamefont {Egelhaaf}}, \bibinfo {author} {\bibfnamefont {M.}~\bibnamefont {Fuchs}}, \bibinfo {author} {\bibfnamefont {J.}~\bibnamefont {Horbach}}, \bibinfo {author} {\bibfnamefont {N.}~\bibnamefont {Koumakis}}, \bibinfo {author} {\bibfnamefont {M.}~\bibnamefont {Kr\"uger}}, \bibinfo {author} {\bibfnamefont {M.}~\bibnamefont {Laurati}}, \bibinfo {author} {\bibfnamefont {K.~J.}\ \bibnamefont {Mutch}}, \bibinfo {author} {\bibfnamefont {G.}~\bibnamefont {Petekidis}}, \bibinfo {author} {\bibfnamefont {M.}~\bibnamefont {Siebenb\"urger}}, \bibinfo {author} {\bibfnamefont {T.}~\bibnamefont {Voigtmann}},\ and\ \bibinfo {author} {\bibfnamefont {J.}~\bibnamefont {Zausch}},\ }\bibfield  {title} {\bibinfo {title} {Residual stresses in glasses},\ }\href {https://doi.org/10.1103/PhysRevLett.110.215701} {\bibfield  {journal}
  {\bibinfo  {journal} {Phys. Rev. Lett.}\ }\textbf {\bibinfo {volume} {110}},\ \bibinfo {pages} {215701} (\bibinfo {year} {2013})}\BibitemShut {NoStop}%
\bibitem [{\citenamefont {Li}\ \emph {et~al.}(2026)\citenamefont {Li}, \citenamefont {Pan}, \citenamefont {Qu},\ and\ \citenamefont {Jin}}]{jin_2026}%
  \BibitemOpen
  \bibfield  {author} {\bibinfo {author} {\bibfnamefont {B.}~\bibnamefont {Li}}, \bibinfo {author} {\bibfnamefont {D.}~\bibnamefont {Pan}}, \bibinfo {author} {\bibfnamefont {T.}~\bibnamefont {Qu}},\ and\ \bibinfo {author} {\bibfnamefont {Y.}~\bibnamefont {Jin}},\ }\bibfield  {title} {\bibinfo {title} {Universal activated aging and weak ergodicity breaking in spin and structural glasses},\ }\href {https://doi.org/10.1126/sciadv.aec4416} {\bibfield  {journal} {\bibinfo  {journal} {Sci. Adv.}\ }\textbf {\bibinfo {volume} {12}},\ \bibinfo {pages} {eaec4416} (\bibinfo {year} {2026})}\BibitemShut {NoStop}%
\bibitem [{IT()}]{IT}%
  \BibitemOpen
  \bibinfo {note} {Coexistence of the Incomplete Thermalization and Mpemba Effect, In preperation}\BibitemShut {NoStop}%
\bibitem [{\citenamefont {Vakarelski}\ \emph {et~al.}(2012)\citenamefont {Vakarelski}, \citenamefont {Patankar}, \citenamefont {Marston}, \citenamefont {Chan},\ and\ \citenamefont {Thoroddsen}}]{vakarelski2012stabilization}%
  \BibitemOpen
  \bibfield  {author} {\bibinfo {author} {\bibfnamefont {I.~U.}\ \bibnamefont {Vakarelski}}, \bibinfo {author} {\bibfnamefont {N.~A.}\ \bibnamefont {Patankar}}, \bibinfo {author} {\bibfnamefont {J.~O.}\ \bibnamefont {Marston}}, \bibinfo {author} {\bibfnamefont {D.~Y.~C.}\ \bibnamefont {Chan}},\ and\ \bibinfo {author} {\bibfnamefont {S.~T.}\ \bibnamefont {Thoroddsen}},\ }\bibfield  {title} {\bibinfo {title} {Stabilization of leidenfrost vapour layer by textured superhydrophobic surfaces},\ }\href {https://doi.org/10.1038/nature11418} {\bibfield  {journal} {\bibinfo  {journal} {Nature}\ }\textbf {\bibinfo {volume} {489}},\ \bibinfo {pages} {274} (\bibinfo {year} {2012})}\BibitemShut {NoStop}%
\bibitem [{\citenamefont {Leidenfrost}(1756)}]{leidenfrost1756de}%
  \BibitemOpen
  \bibfield  {author} {\bibinfo {author} {\bibfnamefont {J.~G.}\ \bibnamefont {Leidenfrost}},\ }\href@noop {} {\emph {\bibinfo {title} {De Aquae Communis Nonnullis Qualitatibus Tractatus}}}\ (\bibinfo  {publisher} {Ovenius},\ \bibinfo {address} {Duisburg},\ \bibinfo {year} {1756})\BibitemShut {NoStop}%
\bibitem [{\citenamefont {Leidenfrost}\ and\ \citenamefont {Wares}(1966)}]{leidenfrost1966translation}%
  \BibitemOpen
  \bibfield  {author} {\bibinfo {author} {\bibfnamefont {J.~G.}\ \bibnamefont {Leidenfrost}}\ and\ \bibinfo {author} {\bibfnamefont {C.}~\bibnamefont {Wares}},\ }\bibfield  {title} {\bibinfo {title} {On the fixation of water in diverse fire},\ }\href {https://doi.org/10.1016/0017-9310(66)90111-6} {\bibfield  {journal} {\bibinfo  {journal} {Int. J. Heat Mass Transfer}\ }\textbf {\bibinfo {volume} {9}},\ \bibinfo {pages} {1153} (\bibinfo {year} {1966})}\BibitemShut {NoStop}%
\bibitem [{\citenamefont {Shirota}\ \emph {et~al.}(2016)\citenamefont {Shirota}, \citenamefont {van Limbeek}, \citenamefont {Sun}, \citenamefont {Prosperetti},\ and\ \citenamefont {Lohse}}]{Leidenfrost}%
  \BibitemOpen
  \bibfield  {author} {\bibinfo {author} {\bibfnamefont {M.}~\bibnamefont {Shirota}}, \bibinfo {author} {\bibfnamefont {M.~A.~J.}\ \bibnamefont {van Limbeek}}, \bibinfo {author} {\bibfnamefont {C.}~\bibnamefont {Sun}}, \bibinfo {author} {\bibfnamefont {A.}~\bibnamefont {Prosperetti}},\ and\ \bibinfo {author} {\bibfnamefont {D.}~\bibnamefont {Lohse}},\ }\bibfield  {title} {\bibinfo {title} {Dynamic leidenfrost effect: Relevant time and length scales},\ }\href {https://doi.org/10.1103/PhysRevLett.116.064501} {\bibfield  {journal} {\bibinfo  {journal} {Phys. Rev. Lett.}\ }\textbf {\bibinfo {volume} {116}},\ \bibinfo {pages} {064501} (\bibinfo {year} {2016})}\BibitemShut {NoStop}%
\end{thebibliography}%

\clearpage
\onecolumngrid{}

\begin{center}
\textbf{\Large{End Matter}}
\end{center}

\twocolumngrid{}

\textsl{\textcolor{black}{Appendix A: Relaxation beyond tight-coupling regime}}\textsl{---}  The theoretical framework presented in the main text naturally generalizes to arbitrary coupling strength ($q \neq 1$). To isolate the specific impact of $q$ on the Mpemba Effect, we consider here the special case of $k_{\alpha}\rightarrow0$. In this case, Eq.~\eqref{eq:Onsager1} and \eqref{eq:Onsager2} reduce to $ J_q=L_{qq}X_q$ and $ J_\alpha=L_{\alpha q}X_q$. Substituting the conjugate thermal force $X_q=\Delta T/T_r^2$ ($\Delta T\equiv T-T_r$) and Eq. \eqref{eq:Onsager2} into these flux equations yields
\begin{equation}
    \frac{\mathrm{d}\Delta T}{\mathrm{d}t}=-\frac{J_q}{{\tilde{C}}}=-\frac{L_{qq}}{\tilde{C}T_r^2}\Delta T,\quad \frac{\mathrm{d}\alpha}{\mathrm{d}t}=L_{\alpha q}\frac{\Delta T}{T_r^2}.
    \label{eq:dTdt}
\end{equation}
By applying the relations $\Delta T=X_qT_r^2=T_r^2J_q/L_{qq}$ and $q=L_{\alpha q}/\sqrt{L_{qq}L_{\alpha\alpha}}$ to $\mathrm{d}\alpha/\mathrm{d}t$ in Eq. \eqref{eq:dTdt}, and subsequently integrating over time, we obtain
\begin{equation}
    \Delta \alpha=\int^t_0 q\sqrt{\frac{L_{\alpha\alpha}}{L_{qq}}}J_q\, \mathrm{d}t'.
    \label{eq:Delta alpha}
\end{equation}
For small structural deviations, this change can be approximated as $\Delta \alpha\approx qk_0 \int_0^tJ_q\,\mathrm{d}t'=qk_0Q(t)$ \citep{Alpha_Q}. Therefore, $L_{qq}=L_{qq}^{(0)}(1+\zeta\Delta\alpha)\approx L_{qq}^{(0)}[1+q\zeta k_0Q(t)]$. Substituting this back into Eq. \eqref{eq:dTdt} directly leads to
\begin{equation}
    \frac{\mathrm{d} T}{\mathrm{d}t} = - \left[ \gamma_0 +q\mathcal{M}Q(t) \right] (T - T_{\mathrm{r}}).
    \label{eq:master_eq_q}
\end{equation}
This represents a generalized form of Eq.~\eqref{eq:master_eq} with $\mathcal{I}=0$. Evidently, the coupling strength effectively rescales the kinetic coefficient $\mathcal{M}$. When heat transfer and structural evolution are uncoupled ($q=0$), Eq.~\eqref{eq:master_eq_q} recovers the conventional Newton's cooling law. As $q$ increases, the cumulative heat transfer integrates more strongly into the relaxation rate, significantly enhancing the ME.

\begin{figure}[htb]
    \centering
    \includegraphics[width=\columnwidth]{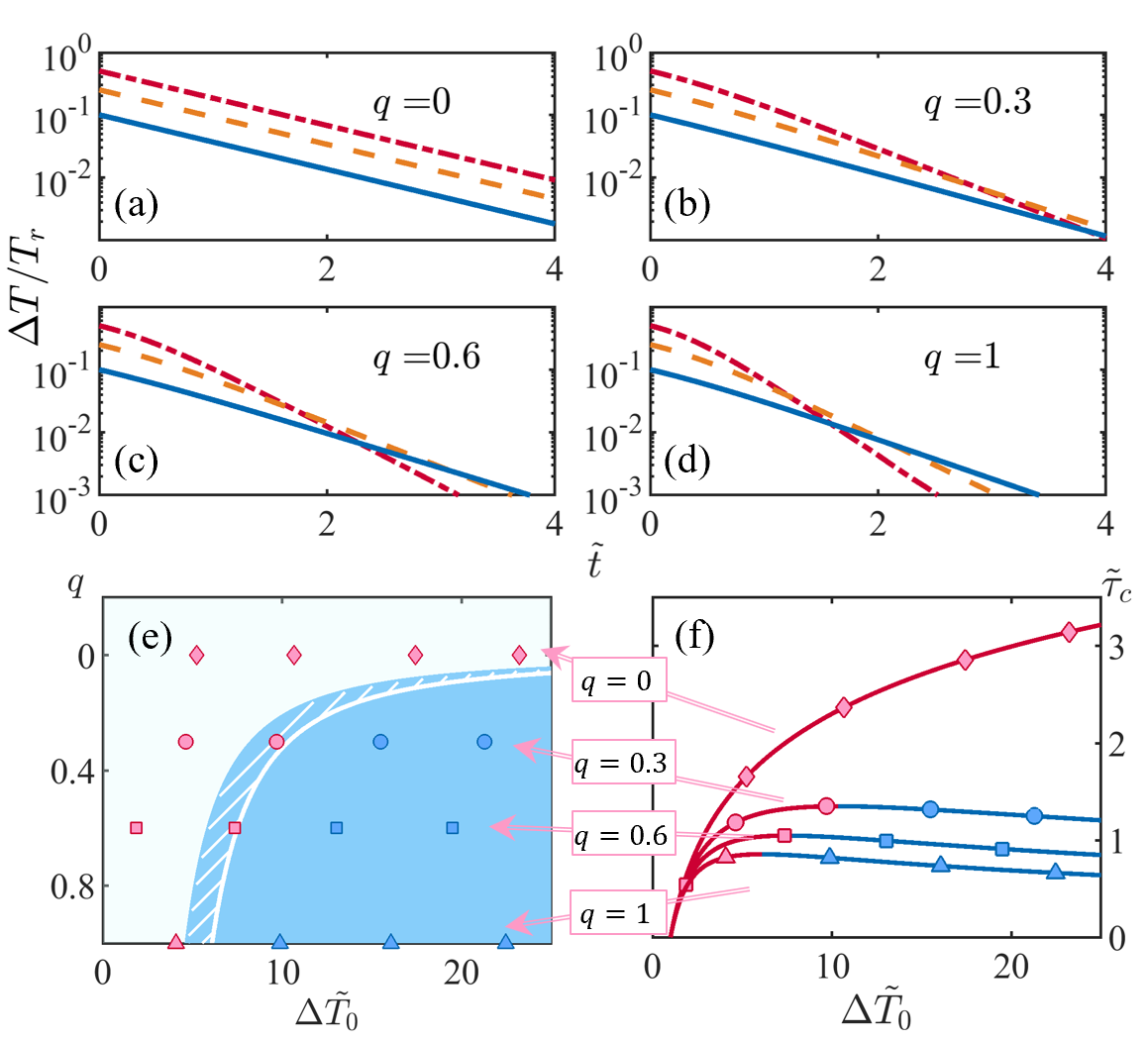}
    \caption{(a-d) $\Delta\tilde{T}=\Delta T/T_r$ as a function of cooling time $\tilde{t}=\gamma_0 t$ with different $q$. The red dash-doted curve, orange dashed curve, and blue solid curve are respectively plotted with $\Delta T_0/T_r=0.5,0.25,0.1$. Other parameters are fixed at $\gamma_0=5\times10^{-3}$, $L_{\alpha\alpha}/L_{qq}^{(0)}=0.01$, $\zeta=0.025$ and $\tilde{C}=10$. (e) Phase diagram in the $(\Delta \tilde{T}_0, q)$ parameter space. The blue area represents the analytically predicted Mpemba effect region, while the white curve denotes the exact numerical boundary. The markers corresponds one-to-one to those on the cooling-time curves plotted in (f). (f) Numerical cooling time $\tilde{\tau}_c$ versus $\Delta \tilde{T}_0=\Delta T_0/\Delta T_f$ for $q=0, 0.3, 0.6, 1$ (top to bottom). Parameters are fixed at $\gamma_0=5\times10^{-3}$, $k_0=0.1$, $\mathcal{M}=5\times10^{-5}$ and $\Delta T_f/T_r=0.025$.}
    \label{fig:T-t with q&phasedia}
    
\end{figure}

Solving Eq.~\eqref{eq:master_eq_q} yields an analytical result for the instantaneous temperature difference
\begin{equation}
    \frac{\Delta T(t)}{\Delta T_0} = \frac{\gamma_0^{-1}q\mathcal{M}\tilde{C}\Delta T_0+1}{\gamma_0^{-1}q\mathcal{M}\tilde{C}\Delta T_0+{e^{(q\mathcal{M}\tilde{C}\Delta T_0+\gamma_0)t}}},
    \label{eq:exact_T_q}
\end{equation}
which is plotted in Fig.~\ref{fig:T-t with q&phasedia}(a-d) across different $q$. When $q=0$, the parallel cooling trajectories in Fig.~\ref{fig:T-t with q&phasedia}(a) indicate the absence of the memory effects. Conversely, any non-zero cross-coupling ($q\neq0$) triggers trajectory crossovers [Fig. \ref{fig:T-t with q&phasedia}(b-d)], which is the direct signature of the ME. Notably, increasing $q$ amplifies the memory response, shifting these intersections to earlier times. Furthermore, the cooling time $\tau_c$, obtained by setting $\Delta T(t)=\Delta T_f$ in Eq.~\eqref{eq:exact_T_q}, recovers Eq.~\eqref{eq:exact_tau} via the rescaling $\mathcal{M} \to q\mathcal{M}$. Evaluating the order parameter $\chi=\partial\tau_{c}/\partial T_{0}$, we obtain the analytical phase diagram in $(\Delta \tilde{T}_0, q)$ space [Fig. \ref{fig:T-t with q&phasedia}(e)], where the ME ($\chi<0$, blue region) and normal cooling ($\chi>0$) are clearly separated. To assess the exactness of this analytical boundary, we perform rigorous numerical simulations for selected states [markers in Fig.~\ref{fig:T-t with q&phasedia}(e)]. The corresponding exact numerical $\tilde{\tau}_c-\Delta \tilde{T}_0$ curves are shown in Fig.~\ref{fig:T-t with q&phasedia}(f), where red and blue segments indicate the increasing and decreasing of $\tilde{\tau}_c$. A cross-examination reveals a minor shift: some markers analytically assigned to the ME region correspond to numerical states that have yet to cross the $\tau_c$ maxima. Tracing the true numerical maxima delineates the exact ME boundary [white curve in Fig.~\ref{fig:T-t with q&phasedia}(e)]. The hatched region, representing the analytical overestimation, stems intrinsically from the small-structural-deviation approximation in Eq.~\eqref{eq:Delta alpha}. Despite this slight boundary shift, our overarching framework remains highly robust in predicting the ME across the parametric space. The details of the numerical simulations are given in Supplemental Materials.

\textsl{\textcolor{black}{Appendix B: The Inverse Mpemba Effect and a Taxonomy of Anomalous Relaxation}}\textsl{---} 
The universal framework naturally encompasses the inverse Mpemba effect (IME), where a system prepared at a lower initial temperature heats up to a target temperature faster than one prepared at a higher initial temperature. During a heating process ($T_0 < T_{\mathrm{r}}$), the system absorbs heat from the reservoir. We define the cumulative heat absorbed by the system as $Q_h(t) = \tilde{C}(T(t) - T_0) = -Q(t) > 0$. Following the biphasic reservoir model discussed in Example I, the reservoir now releases heat, and the time-dependent thermal conductance is modified as $\Gamma(t) = \Gamma_0 - \Theta^{-1}(\mu-1)\Gamma_2 Q_h(t)$.
Using the memory response coefficient $\mathcal{M} = (\mu-1)\Gamma_2 / (\Theta \tilde{C})$, the exact heating dynamics become
\begin{equation}
    \frac{\mathrm{d}T}{\mathrm{d}t} = \left[ \gamma_0 - \mathcal{M}\tilde{C}(T - T_0) \right] (T_{\mathrm{r}} - T). \label{eq:B1}
\end{equation}
To trigger the IME, the dynamic conductance must be enhanced as the system absorbs heat. This requires the newly formed phase (induced by the reservoir releasing heat) to have a higher thermal conductance. Since the reservoir releases heat to the system, phase $\Phi_1$ turns into phase $\Phi_2$. Therefore, we must have $\Gamma_2 > \Gamma_1$, which corresponds to $\mu \equiv \Gamma_1/\Gamma_2 < 1$. Consequently, $\mathcal{M} < 0$. Let $\mathcal{M}' = -\mathcal{M} > 0$. The effective heating rate is $\gamma_{\text{eff}}(t) = \gamma_0 + \mathcal{M}' \tilde{C}(T - T_0)$. A system starting from a lower initial temperature $T_0$ (i.e., further away from the target $T_{\mathrm{r}}$) will undergo a larger temperature change $(T - T_0)$ to reach a given intermediate target $T$. This induces a dynamically larger effective heating rate (since $-\mathcal{M}Q > 0$), mathematically enabling the colder trajectory to sustain a higher velocity and overtake the warmer one. This strictly guarantees the occurrence of the IME, perfectly consistent with the numerical results depicted in Fig.~\ref{fig:Temperature-different}(d).

This profound directional duality, where the exact same negative memory ($\mathcal{M} < 0$) acts as a powerful accelerator during heating but as a drastic brake during cooling, reveals that the predictive power of Eq.~(\ref{eq:master_eq}) extends far beyond the ME and IME. By systematically analyzing the interplay between the thermodynamic process (direction of heat flow) and the sign of $\mathcal{M}$, we establish a complete $2 \times 2$ macroscopic taxonomy of memory-dependent anomalous relaxation, as summarized in Table~\ref{tab:taxonomy}.

\begin{table}
\centering
\caption{\label{tab:taxonomy}Taxonomy of memory-dependent anomalous relaxation. The dynamic evolution of the effective relaxation rate $\gamma_{\text{eff}}$ ($\uparrow$: enhanced, $\downarrow$: suppressed) dictates the emergence of diverse anomalous phenomena.}
\renewcommand{\arraystretch}{1.5}
\begin{tabular*}{8.5cm}{@{\extracolsep{\fill}}lccc@{}}
\toprule
\toprule
\textbf{Process} & $\boldsymbol{\mathcal{M}}$ & $\boldsymbol{\gamma_{\text{eff}}}$ & \textbf{Phenomenon} \\
\midrule
\multirow{2}{*}{\shortstack[l]{Cooling \\ ($T_0 > T_{\mathrm{r}}$)}} & $\mathcal{M} > 0$ & $\uparrow$ & ME \\
 & $\mathcal{M} < 0$ & $\downarrow$ & Anti-ME \\
\midrule
\multirow{2}{*}{\shortstack[l]{Heating \\ ($T_0 < T_{\mathrm{r}}$)}} & $\mathcal{M} < 0$ & $\uparrow$ & Inverse ME \\
 & $\mathcal{M} > 0$ & $\downarrow$ & Anti-Inverse ME \\
\bottomrule
\bottomrule
\end{tabular*}
\end{table}

The effective relaxation rate is universally governed by $\gamma_{\text{eff}} = \gamma_0 + \mathcal{M}Q(t)$. During a cooling process ($Q > 0$), a positive memory ($\mathcal{M} > 0$) dynamically enhances heat transfer, recovering the forward ME. Conversely, if $\mathcal{M} < 0$, the dynamically generated structure severely hinders heat transfer, suppressing the cooling of a hotter initial state and triggering the Anti-ME, as noted in the main text. During a heating process ($Q < 0$), the dynamics invert. As explicitly derived above, $\mathcal{M} < 0$ acts as an accelerator, rigorously guaranteeing the IME. Finally, the framework completes the taxonomy by predicting the Anti-Inverse ME for heating processes with $\mathcal{M} > 0$, where the thermalization of a colder initial state is anomalously retarded. A classic realization of this Anti-IME is the traditional Leidenfrost effect observed when a cold water drop is placed on a scorching hot surface \citep{leidenfrost1756de,leidenfrost1966translation,Leidenfrost}: the drop rapidly evaporates to form a stable vapor cushion, drastically suppressing the heating rate and allowing the droplet to persist.
\end{document}


\begin{CJK*}{UTF8}{gbsn}

\title{Supplemental Materials for: \\ ``Macroscopic Thermodynamic Framework for the Mpemba Effect''}       
\author{Yun-Qian Lin (林蕴芊)}
\affiliation{School of Physics and Astronomy, Beijing Normal University, Beijing, 100875, China}

\author{Z. C. Tu (涂展春)}
\address{School of Physics and Astronomy, Beijing Normal University, Beijing, 100875, China}
\address{Key Laboratory of Multiscale Spin Physics (Beijing Normal University), Ministry of Education, Beijing 100875, China}

\author{Yu-Han Ma (马宇翰)}
\email{yhma@bnu.edu.cn}
\address{School of Physics and Astronomy, Beijing Normal University, Beijing, 100875, China}
\address{Key Laboratory of Multiscale Spin Physics (Beijing Normal University), Ministry of Education, Beijing 100875, China}
\date{\today}

\maketitle
\end{CJK*}
\onecolumngrid
\begin{spacing}{1.7} 
\textbf{This document is devoted to providing the detailed derivations and the supporting discussions to the main text of the Letter. The contents of the Supplemental Materials are listed as follows}

\tableofcontents

\end{spacing}
\newpage

\setlength{\parskip}{1em}
    \section{Generality of the Main Framework}

\subsection{Multi-dimensional Structural Variables}
The theoretical framework presented in the main text naturally generalizes to complex systems governed by multiple macroscopic slow variables. This extension is crucial for realistic setups where the memory carrier possesses a complex energy landscape, multiple internal structural modes, or spatially distributed boundary interactions. We consider a set of $M$ macroscopic generalized variables denoted by the vector $\boldsymbol{\alpha}(t) = \{\alpha_1(t), \alpha_2(t), \dots, \alpha_M(t)\}$. The internal energy is generalized as $E = E(T, \boldsymbol{\alpha})$. The total entropy of the isolated composite system is now parameterized as $S_{\mathrm{tot}}(E, \alpha_1, \dots, \alpha_M)$, then the total entropy production rate reads
\begin{equation}
    \sigma = \frac{\partial S_{\mathrm{tot}}}{\partial E} \frac{\mathrm{d} E}{\mathrm{d} t} + \sum_{i=1}^M \frac{\partial S_{\mathrm{tot}}}{\partial \alpha_i} \frac{\mathrm{d} \alpha_i}{\mathrm{d} t}. \label{eq:entropy_rate_M}
\end{equation}
The heat flux released by the system is given by the total energy decrease 
\begin{equation}
J_q = -\frac{\mathrm{d}E}{\mathrm{d}t} = -C\frac{\mathrm{d}T}{\mathrm{d}t} - \sum_{i=1}^M u_i J_{\alpha_i},   
\end{equation}
where $C = \partial E/\partial T$, $u_i = \partial E/\partial \alpha_i$ represents the generalized internal latent heat associated with the $i$-th structural mode, and $J_{\alpha_i} = \mathrm{d}\alpha_i/\mathrm{d}t$ denotes the structural flux. The bilinear form of the entropy production is then 
\begin{equation}
    \sigma = J_q X_q + \sum_i J_{\alpha_i} X_{\alpha_i} \ge 0.
\end{equation}
The conjugate thermal force remains $X_q \approx (T-T_{\mathrm{r}})/T_{\mathrm{r}}^2$. For the $M$-dimensional generalized restorative forces, we expand the entropy around equilibrium. By relating the entropy change to the generalized free energy $\Delta S_{\mathrm{tot}} = -\Delta F / T_{\mathrm{r}}$, we introduce a symmetric, positive-definite generalized stiffness matrix $K_{ij} = \partial^2 F / (\partial \alpha_i \partial \alpha_j)$. Therefore, the restoring forces read
\begin{equation}
X_{\alpha_i}= \frac{\partial S_{\mathrm{tot}}}{\partial \alpha_i} \approx -\frac{1}{T_{\mathrm{r}}} \sum_{j=1}^M K_{ij} \Delta \alpha_j. \label{eq:X_alpha_M}
\end{equation}

In this generalized case, the macroscopic dynamics are governed by the $(M+1) \times (M+1)$ Onsager matrix as
\begin{align}
    J_q &= L_{qq} X_q + \sum_{i=1}^M L_{q \alpha_i} X_{\alpha_i}, \label{eq:Onsager1_M} \\
    J_{\alpha_i} &= L_{\alpha_i q} X_q + \sum_{j=1}^M L_{\alpha_i \alpha_j} X_{\alpha_j}. \label{eq:Onsager2_M}
\end{align}
We assume the system operates in the global \textit{tight-coupling limit}, where $q=L_{\alpha_iq}/\sqrt{L_{\alpha_i\alpha_i}L_{qq}}=1$. Physically, this means all structural fluxes are strictly proportional to the primary thermal flux, i.e.,$J_{\alpha_i} = k_i J_q$, where $k_i \equiv \sqrt{L_{\alpha_i\alpha_i} / L_{qq}}=L_{\alpha_iq}/L_{qq}$ represents the specific kinematic coupling constant for the $i$-th structural mode. Substituting this tight-coupling constraint back into the heat flux decomposition $J_q = -C\mathrm{d}T/\mathrm{d}t - \sum_i u_i J_{\alpha_i}$, we obtain 
\begin{equation}
    J_q = -C\mathrm{d}T/\mathrm{d}t - (\sum_i k_i u_i) J_q.
\end{equation}
Rearranging this isolates the temperature evolution rate
\begin{equation}
    C \frac{\mathrm{d}T}{\mathrm{d}t} = -J_q \left( 1 + \sum_{i=1}^M k_i u_i \right) \equiv -J_q \frac{C}{\tilde{C}}. \label{eq:renormalized_C}
\end{equation}
Here, the internal latent heat dynamics are absorbed by defining a generalized effective heat capacity $\tilde{C} = C / (1 + \sum_{i=1}^M k_i u_i)$,  where we assume $1 + \sum_i k_i u_i > 0$ to ensure thermodynamic stability ($\tilde{C}>0$). Meanwhile, substituting $L_{q \alpha_i} = k_i L_{qq}$ (due to Onsager reciprocity $L_{q \alpha_i} = L_{\alpha_i q}$) back into Eq.~(\ref{eq:Onsager1_M}), the heat flux is expressed as
\begin{equation}
    J_q = L_{qq} \left( X_q + \sum_{i=1}^M k_i X_{\alpha_i} \right). \label{eq:Jq_factored_M}
\end{equation}

Next, we consider the dynamic response of the principal transport coefficient $L_{qq}$ to the multidimensional structural deformations. To the leading order, we expand it as
\begin{equation}
 L_{qq}(\boldsymbol{\alpha}) \approx L_{qq}^{(0)} \left(1+\sum_{i=1}^M \zeta_i\Delta \alpha_i\right),
\end{equation}
where $\zeta_i \equiv (\partial L_{qq}/\partial \alpha_i)/L_{qq}^{(0)}$ quantifies the sensitivity to the $i$-th structural mode. To track the structural evolution, we substitute the definition of the coupling constant $k_i(\boldsymbol{\alpha}) = \sqrt{L_{\alpha_i \alpha_i}/L_{qq}(\boldsymbol{\alpha})}$ into the exact kinetic relation $\mathrm{d}\alpha_i = k_i(\boldsymbol{\alpha}) \mathrm{d}Q$. Expanding this yields
\begin{equation}
\frac{\mathrm{d}\alpha_i}{\mathrm{d}Q} = \frac{\sqrt{L_{\alpha_i \alpha_i}}}{\sqrt{L_{qq}^{(0)} \left(1+\sum_{j=1}^M \zeta_j\Delta \alpha_j\right)}} = k_i^{(0)} \left[ 1 - \frac{1}{2} \sum_{j=1}^M \zeta_j \Delta\alpha_j + \mathcal{O}(\Delta\boldsymbol{\alpha}^2) \right],
\end{equation}
where $k_i^{(0)} \equiv \sqrt{L_{\alpha_i \alpha_i}/L_{qq}^{(0)}}$ is the unperturbed coupling constant. Within the small structural deviation regime ($|\zeta_j \Delta\alpha_j| \ll 1$), the higher-order structural feedback on the kinematic coupling can be neglected. Retaining only the leading order, the above differential relation simplifies to $\mathrm{d}\alpha_i \approx k_i^{(0)} \mathrm{d}Q$, which integrates directly as
\begin{equation}
\Delta \alpha_i(t) \approx \int_0^t k_i^{(0)} J_q \mathrm{d}t' = k_i^{(0)} Q(t).
\label{eq:alpha_Q}
\end{equation}
Substituting Eq.~(\ref{eq:alpha_Q}) into the expansion of $L_{qq}$ yields the memory-dependent transport coefficient
\begin{equation}
L_{qq}(Q)  \approx L_{qq}^{(0)} \left[ 1+\left(\sum_{i=1}^M k_i^{(0)}\zeta_i\right)Q(t) \right]. \label{eq:Lqq_M}
\end{equation}
Simultaneously, we evaluate the combined driving force in Eq.~(\ref{eq:Jq_factored_M}). Substituting Eqs.~(\ref{eq:X_alpha_M}) and (\ref{eq:alpha_Q}) along with $X_q$, we obtain
\begin{equation}
X_q + \sum_{i=1}^M k_i^{(0)} X_{\alpha_i} = \frac{T - T_{\mathrm{r}}}{T_{\mathrm{r}}^2} - \frac{1}{T_{\mathrm{r}}} \left( \sum_{i=1}^M \sum_{j=1}^M K_{ij}k_i^{(0)}k_j^{(0)} \right) Q(t). \label{eq:Forces_M}
\end{equation}
Finally, substituting Eqs.~(\ref{eq:Lqq_M}) and (\ref{eq:Forces_M}) into Eq.~(\ref{eq:Jq_factored_M}) and utilizing the renormalized relation $J_q = -\tilde{C} \mathrm{d}T/\mathrm{d}t$, we arrive at the fully generalized thermal relaxation equation
\begin{equation}
\frac{\mathrm{d}T}{\mathrm{d}t} = - \left[ \gamma_0 + \mathcal{M}Q(t) \right] \big[ (T - T{\mathrm{r}}) - \mathcal{I}Q(t) \big]. \label{eq:master_eq_M}
\end{equation}
Here, the initial base rate is given by $\gamma_0 = L_{qq}^{(0)}/(\tilde{C} T_{\mathrm{r}}^2)$, and the effective memory response coefficient $\mathcal{M}$ and the structural inertia coefficient $\mathcal{I}$ now embed the full complexity of the multidimensional phase space and internal energy coupling
\begin{equation}
    \mathcal{M}= \gamma_0\sum_{i=1}^Mk_i^{(0)}\zeta_i,~~ \mathcal{I}= T_{\mathrm{r}} \sum_{i=1}^M \sum_{j=1}^M K_{ij} k_i^{(0)}k_j^{(0)}.
\end{equation}
Equation~(\ref{eq:master_eq_M}) reveals a profound structural invariance. Even when the physical system possesses arbitrarily complex multi-mode dynamics and intrinsic energy-structure couplings (latent heats), as long as it operates under the tight-coupling constraint, the macroscopic temperature evolution strictly collapses onto the exact same single-variable functional form as [Eq.~(6)] derived in the main text.

\subsection{Numerical Analysis of the Relaxation beyond Tight-coupling Regime}
In Appendix A of the End matter of the main text, the influence of coupling strength $q$ on the ME in the limit of $k_{\alpha}\to 0$ has been analytically discussed under the approximation of small structural deviation. Here, we provide the direct numerical calculations and examine the consistency of the analytical and numerical results.

Eliminating the time variable in Eq.~(6) in Appendix A and combining the first-order expansion of $L_{qq}$ with respect to $\Delta\alpha$, yields the differential relation between the temperature deviation and the structural variable
\begin{equation}
    \frac{\mathrm{d}\Delta T}{\mathrm{d}\Delta \alpha}=-\frac{L_{qq}^{(0)}(1+\zeta\Delta\alpha)}{\tilde{C}L_{\alpha q}}.
    \label{eq:dT-dalpha}
\end{equation}
where the cross-coupling coefficient can be expressed as $L_{\alpha q}=q\sqrt{L_{\alpha \alpha }L_{qq}^{(0)}(1+\zeta\Delta \alpha)}$. By integrating Eq.~\eqref{eq:dT-dalpha} in the coupled region $(q\neq0)$, we obtain the temperature difference as a function of the structural deviation
\begin{equation}
    \Delta T-\Delta T_0=-\frac{2 \left[\left( L_{qq}^{(0)}(1+\zeta\Delta\alpha)\right)^{3/2}-{L_{qq}^{(0)}}^{3/2}\right]}{3q\tilde{C}\zeta L_{qq}^{(0)}\sqrt{L_{\alpha \alpha }}}.
    \label{eq:Delta T-alpha}
\end{equation}
Substituting Eq.~\eqref{eq:Delta T-alpha} into $\mathrm{d}\alpha/\mathrm{d}t$ in [Eq.~(16)] leads to the evolution equation of the structural variable
\begin{equation}
    \frac{\mathrm{d}\alpha }{\mathrm{d}t}=\frac{2L_{\alpha q}}{T_r^2}\left\{\frac{\Delta T_0}{2}-\frac{\left[L_{qq}^{(0)}(1+\zeta\Delta\alpha)\right]^{3/2}-\left({L_{qq}^{(0)}}\right)^{3/2}}{3q\tilde{C}\zeta L_{qq}^{(0)}\sqrt{L_{\alpha \alpha }}}\right\},
    \label{eq:d alpha/dt-alpha}
\end{equation}
Eq.~\eqref{eq:d alpha/dt-alpha} explicitly indicates that any non-zero cross coupling introduces a positive correlation between the structural relaxation rate and the initial temperature, which in turn affects the temperature evolution. Numerically solving Eq.~\eqref{eq:d alpha/dt-alpha} and substituting the results into Eq.~\eqref{eq:Delta T-alpha} , we obtain the temperature evolution. The numerical cooling trajectories show the behavior consistent with that in Appendix A: for different initial temperatures, a stronger coupling shifts the intersections of the trajectories to earlier times. 

\begin{figure}
    \centering
    \includegraphics[width=0.55\linewidth]{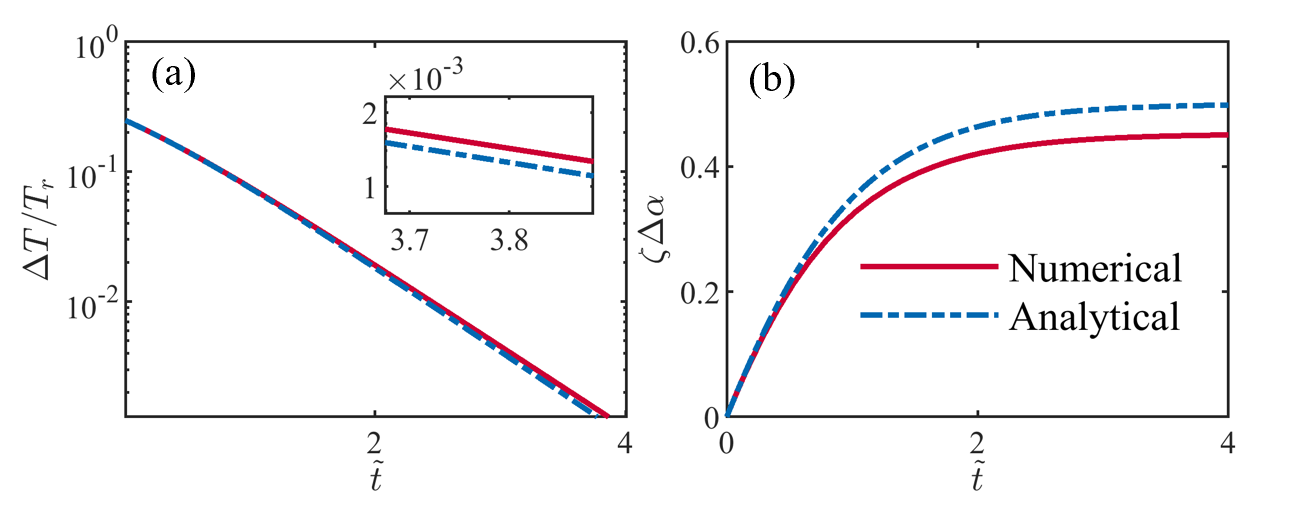}
    \caption{The evolution of temperature and the structural variable against the cooling time $\tilde{t}=\gamma_0 t$ with $q=0.5$.  The red solid lines illustrate the numerical results obtained by solving Eq.~\eqref{eq:d alpha/dt-alpha}, while the blue dash-dotted line denotes the analytical results in Appendix A. Other parameters are fixed at $T_0/T_r=0.25$, $\gamma _0=0.005$, $L_{\alpha\alpha}/L_{qq}^{(0)}=0.01$, $\zeta=0.025$ and $\tilde{C}=10$.}
    \label{fig:consistency of T alpha}
\end{figure}

To further examine the agreement between the analytical approximation and the direct numerical result , we compare the results for a representative initial temperature. Figure \ref{fig:consistency of T alpha} (a) shows the trajectories of temperature as a function of cooling time $\tilde{t}$ with $q=0.5$. The analytical and numerical curves agree well  at the initial stage. As the structural variable grows [Fig.~\ref{fig:consistency of T alpha} (b)], a small deviation emerges. This discrepancy originates from the linear approximation of $L_{qq}$ with respect to $\Delta \alpha$, rather than from the basic assumptions of the main framework.  Therefore, the main framework robustly captures the mechanism of the initial-state- dependent anomalous relaxation.

\section{Inverse Mpemba Effect}
\subsection{Inverse ME beyond Tight-coupling Regime}
\begin{figure}
     \centering
     \includegraphics[width=\linewidth]{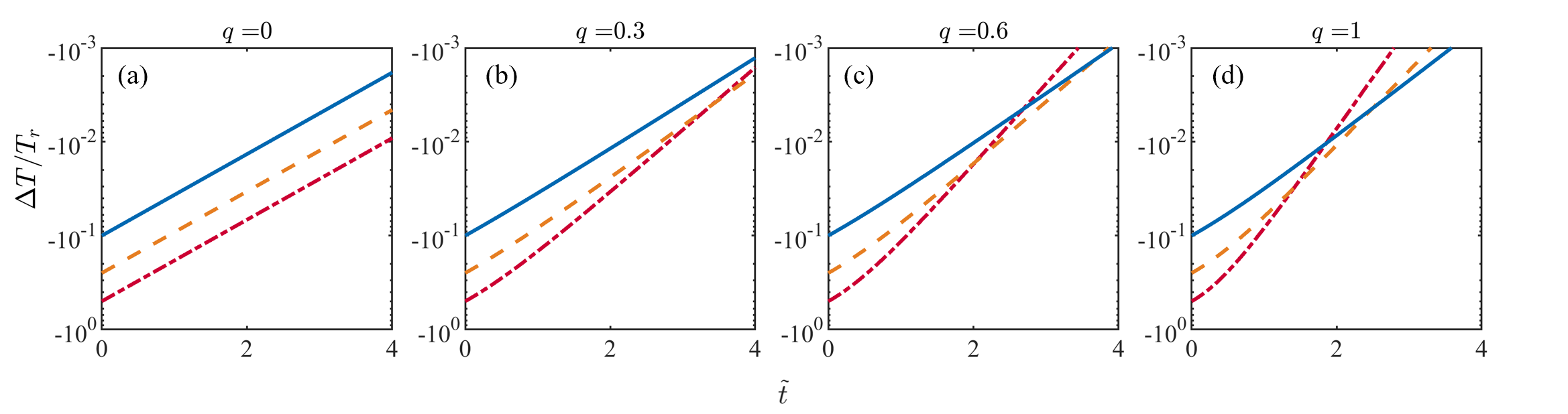}
     \caption{$\Delta\tilde{T}=\Delta T/T_r$ as a function of heating time $\tilde{t}=\gamma_0 t$ with different $q$. The red dash-doted curve, orange dashed curve, and blue solid curve are respectively plotted with $\Delta T_0/T_r=-0.5,-0.25,-0.1$. Other parameters are fixed at $\gamma_0=5\times10^{-3}$, $L_{\alpha\alpha}/L_{qq}^{(0)}=0.01$, $\zeta=-0.025$ and $\tilde{C}=10$.}\label{fig:inverseME-q}
 \end{figure}

 The temperature evolution equation derived in the main text is applicable to both the cooling and heating processes. When $\mathcal{M}<0$, the universal framework naturally allows for the inverse Mpemba effect (IME), in which a system prepared at a lower initial temperature heats up to a target temperature faster than one prepared at a higher initial temperature. The generalized form of the temperature evolution regarding the coupling strength $q$ is exactly [Eq.~(18)] in Appendix A. As $q$ increases, the cumulative heat transfer integrates more strongly into the relaxation rate, significantly enhancing the IME. The temperature difference of heating process is governed by
 \begin{equation}
    \frac{\Delta T(t)}{\Delta T_0} = \frac{\gamma_0^{-1}q\mathcal{M}C\Delta T_0+1}{\gamma_0^{-1}q\mathcal{M}C\Delta T_0+{e^{(q\mathcal{M}C\Delta T_0+\gamma_0)t}}}.
    \label{eq:exact_T_q}
\end{equation}
where $\gamma_0=L_{qq}^{(0)}/CT_r^2$, $\mathcal{M}=\gamma_0k_0\zeta$ and $\zeta=(\partial L_{qq}/\partial \alpha)/L_{qq}^{(0)}$. Fig.~\ref{fig:inverseME-q} plots Eq.~\eqref{eq:exact_T_q} with different $q$, showing that any non-zero cross-coupling $(q\neq0)$ triggers trajectory crossovers [Fig.~\ref{fig:inverseME-q}(b-d)], which are seen as direct evidence of the inverse ME, as shown for ME in Appendix A. Consistently, increasing $q$ amplifies the memory response, shifting these intersections to earlier times.
\subsection{Inverse ME in Example I}
\begin{figure}
    \centering
    \includegraphics[width=0.6\linewidth]{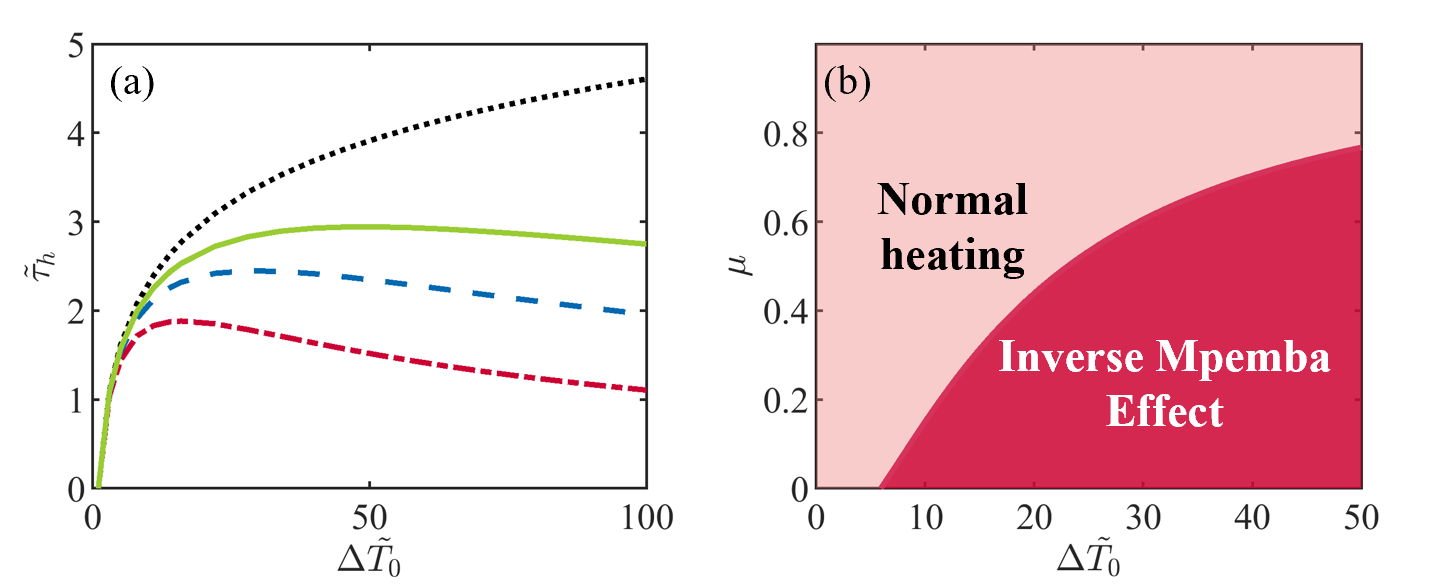}
    \caption{The occurrence of Inverse ME. (a) The heating time $\tilde{\tau}_h\equiv \gamma_0\tau_h$ as a function of dimensionless temperature difference $\Delta\tilde{T_0}\equiv\Delta T_0/\Delta T_f$ with $\mu=1, 0.5, 0.3, 0.1$ (from top to bottom). (b) Phase diagram of the heating process in the parametric space $(\Delta\tilde{T}_0,\mu)$. The inverse ME region is characterized by $\partial\tau_{h}/\partial \Delta\tilde{T}_0<0$ while the normal cooling is associated with $\partial\tau_{h}/\partial \Delta\tilde{T}_0\geq0$}
    \label{fig:inverse ME}
\end{figure}

Following the model discussed in Example I regarding the tight-coupling regime, the exact heating time $\tau_h$ to reach a target temperature takes the identical expression of the cooling time, i.e.,
\begin{equation}
\tau_h(\Delta T_0) = \frac{\ln \left\{\frac{\Delta T_0}{\Delta T_f} \left[1 + \mathcal{M}C \gamma_0^{-1}(\Delta T_0 - \Delta T_f)\right] \right\}}{\gamma_0 + \mathcal{M}C \Delta T_0}, \label{eq:exact_tau}
\end{equation}
where $\gamma_0=\Gamma_0/C$ and $\mathcal{M}=(\mu-1)\Gamma_2/(C\Theta)$. Eq.~\eqref{eq:exact_tau} is plotted in Fig.~\ref{fig:inverse ME} (a) with several values of $\mu$. Notably, for Inverse ME, we must have $\mu<1$ then $\mathcal{M}<0$, as  $\Delta T_0<\Delta T_f$ in the heating process. When $\mu=1$, as the black dotted line illustrated in Fig.~\ref{fig:inverse ME}, the heating time is monotonically increasing with the dimensionless temperature difference. As $\mu$ decreases, $\mathcal{M}$ becomes more negative, corresponding to a stronger kinetic response, which enhances the heating rate. Consequently, an initially colder system can reach the target temperature in a shorter time than an initially warmer one, signaling the emergence of the inverse ME.  The occurrence of Inverse ME is identified by the sign of order parameter $\chi_h\equiv\partial\tau_h/\partial \Delta\tilde{T}_0$: $\chi_h>0$ corresponds to normal heating process, $\chi_h<0$ indicates the region of Inverse ME, as shown in Fig.~\ref{fig:inverse ME}.

\section{Supplementary Analysis of Example I}
\label{Example I}
\begin{figure}[htb]
    \centering
    \includegraphics[width=\columnwidth]{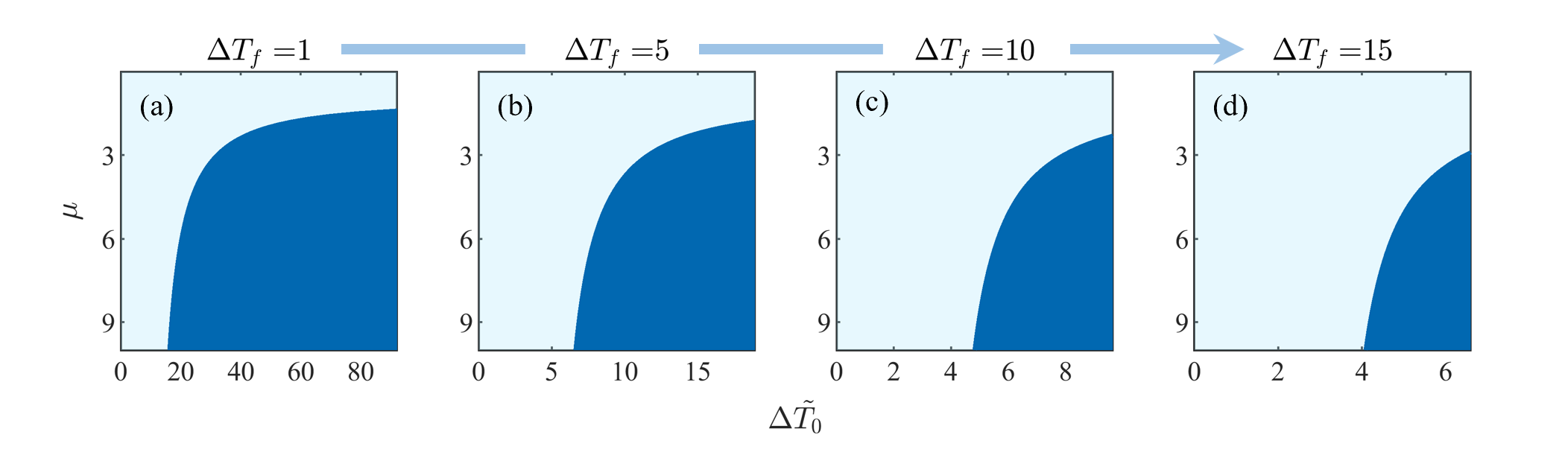}
    \caption{Phase diagram of the cooling process in the parametric space with increasing target final temperature $\Delta T_f$. The blue region denotes the ME area where $\partial \tau_c/\partial T_0<0$.}
    \label{fig:phasedia}
\end{figure} 
In the main text, the phase diagram in the parameter space $(\mu,\Delta \tilde{T}_0)$ is presented for a representative target temperature $\Delta T_f$. Since the cooling time is defined as the time at which $\Delta T(t)$ reaches $\Delta T_f$, the choice of the target temperature can affect the behavior of the phase diagram. As shown in Figure~\ref{fig:phasedia}, increasing $\Delta T_f$ shrinks the region of the ME where the order parameter $\chi=\partial \tau_c/\partial T_0<0$. A larger $\Delta T_f$ corresponds to probing at the early stage of the relaxation, where the heat transfer has not accumulated and the structural response remains weak; hence, the ME region shrinks as $\Delta T_f$ increases.

The ME appears when $\tau_c$ becomes non-monotonic with the initial temperature, i. e., when the cooling time increases initially and decreases afterwards. This behavior indicates the existence of the maximum of cooling time $\tau_{max}$ and the corresponding $\Delta T_0^*$, tracing which across parameter space delineates the exact ME phase boundary. According to Fig.~\ref{fig:phasedia}, the slope of the phase boundary changes with varying $\mu$, which enters the relaxation through the kinetic response coefficient $\mathcal{M}$. In addition, the target final temperature $\Delta T_f$ also modifies the behavior of the phase diagram. To understand how these parameters reshape the boundary, we next analyze the behavior of the maximum point $\Delta T_0^*$, especially its dependence on $\mathcal{M}$ and $\Delta T_f$.

The maximum point $\Delta T_0^*$ is determined by the extremum condition of the cooling time, $\partial\tau_c/\partial T_0|_{\Delta T_0^*}=0$. Substituting [Eq.~(12)] into this condition yields an implicit equation for $\Delta T_0^*$
\begin{equation}
    (1+\mathcal{T}_0)\left[\frac{1}{\mathcal{T}_0}+\frac{1}{1+\mathcal{T}_0-\mathcal{T}_f}\right]=\ln{\frac{\mathcal{T}_0(1+\mathcal{T}_0-\mathcal{T}_f)}{\mathcal{T}_f}},
\label{eq:maximum condition}
\end{equation}
where $\mathcal{T}_0=\mathcal{M}C\Delta T_0^*/\gamma_0$, $\mathcal{T}_f=\mathcal{M}C\Delta T_f/\gamma_0$. Eq~\eqref{eq:maximum condition} indicates that the kinetic response coefficient $\mathcal{M}$ and the target final temperature difference $\Delta T_f$ jointly control the location of  $\Delta T_0^*$ through $\mathcal{T}_f$.

\begin{figure}
    \centering
    \includegraphics[width=0.6\linewidth]{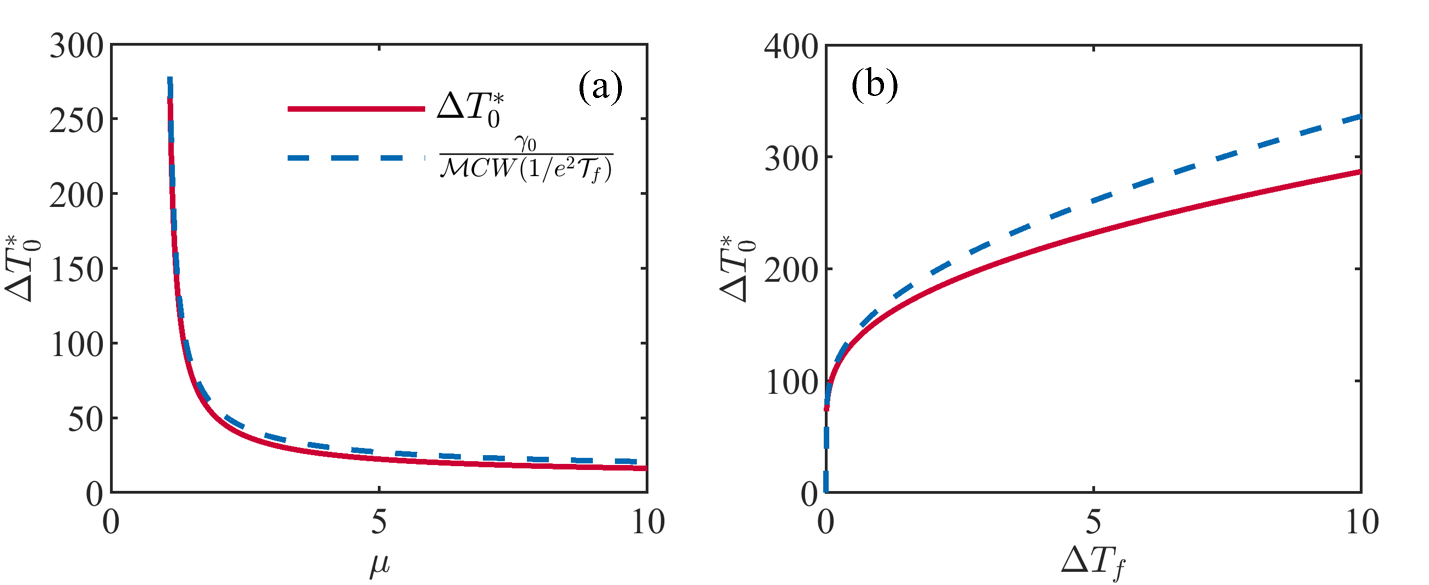}
    \caption{The behavior of the location of maximum cooling time in the regime of $\mathcal{M}\Delta T_f\ll\gamma_0/C$. The red solid lines illustrates the numerical solution of the extremum condition Eq.~\eqref{eq:maximum condition},  and the blue dashed lines denote Eq.~\eqref{eq:T0*-0}. (a) The dependence of $\Delta T_0^*$ on $\mu$ with $\Delta T_f=1$. (b) The dependence of $\Delta T_0^*$ on $\Delta T_f$ with $\mu=1.2$. Other parameters are fixed at $C=1$, $n_0=0.1$, $\Gamma_2=5$ and $\Theta=100$.}
    \label{fig:T0*-0}
\end{figure}

We first consider the limit of $\mathcal{M}\Delta T_f \ll \gamma_0/C$, or equivalently, $\mathcal{T}_f\ll 1$. In this regime, Eq.~\eqref{eq:maximum condition} reduces to 
\begin{equation}
    2+\frac{1}{\mathcal{T}_0} \simeq
    \ln
    \left[
    \frac{\mathcal{T}_0(1+\mathcal{T}_0)}
    {\mathcal{T}_f}
    \right] \label{eq:maximumT0-0}
\end{equation}
The right-hand side contains a large contribution of the order $\ln(1/\mathcal{T}_f)$. If $\mathcal{T}_0=O(1)$, the left-hand side remains finite, while the right-hand side diverges as $\mathcal{T}_f\to0$.  The possibility $\mathcal{T}_0\gg1$ is also excluded. In the regime of $\mathcal{T}_f\ll 1$, the left-hand side approaches a finite value ($2+1/\mathcal{T}_0\simeq 2$), whereas the right-hand side $\simeq\ln[(\mathcal{T}_0)^2/\mathcal{T}_f]$ diverges. 
The only remaining possibility is that $\mathcal{T}_0 \ll 1$. Meanwhile, since the cooling time is evaluated for $\Delta T_0>\Delta T_f$, the self-consistent scaling regime of $\mathcal{T}_0$ is $\mathcal{T}_f \ll \mathcal{T}_0 \ll 1 $. Hence, Eq.~\eqref{eq:maximumT0-0} becomes
\begin{equation}
    \frac{1}{\mathcal{T}_0}+2
    \simeq
    \ln\frac{\mathcal{T}_0}{\mathcal{T}_f}.
    \label{eq:maximu condition-0}
\end{equation}
Substituting the definitions of $\mathcal{T}_0$ and $\mathcal{T}_f$ back into Eq.~\eqref{eq:maximu condition-0} leads to
\begin{equation}
    \Delta T_0^*
    \simeq
    \frac{\gamma_0}{\mathcal{M}C}
    \frac{1}{
    W\left(
    \dfrac{\gamma_0}
    {e^2\mathcal{M}C\Delta T_f}
    \right)
    }\sim 
    \frac{\gamma_0}{\mathcal{M}C}
    \left[\ln\left(
    \dfrac{\gamma_0}
    {\mathcal{M}C\Delta T_f}
    \right)\right]^{-1},
    \label{eq:T0*-0}
\end{equation}
where $W$ denotes the Lambert $W$ function, defined by $W(z)e^{W(z)}=z$. 
Eq.~\eqref{eq:T0*-0} shows the dependence of the maximum point on $\mathcal{M}$ and $\Delta T_f$ in the regime of $\mathcal{M}\Delta T_f\ll\gamma_0/C$.  It reveals that increasing $\mathcal{M}$ shifts the maximum point to a smaller $\Delta T_0^*$, approximately scaling as $1/\mathcal{M}$ with a logarithmic correction; hence, the phase boundary changes rapidly with $\mathcal{M}$ in this regime. In Example I, since $\mathcal{M}=(\mu-1)\Gamma_2/\Theta C$, $\Delta T_0^*$ exhibits the same qualitative dependence on $\mu$ as on $\mathcal{M}$. Fig.~\ref{fig:T0*-0} (a) plots Eq.~\eqref{eq:T0*-0} together with the direct numerical solutions of Eq.~\eqref{eq:maximum condition} as a function of $\mu$, showing that the approximate expression captures the numerical result well within the regime of $\mathcal{M}\Delta T_f\ll\gamma_0/C$.
Conversely, increasing $\Delta T_f$ leads to a larger $\Delta T_0^*$. The positive dependence on $\Delta T_f$ enters only through the logarithmic factor and is therefore relatively weak, as shown in Fig.~\ref{fig:T0*-0} (b). 

\begin{figure}
    \centering
    \includegraphics[width=0.6\linewidth]{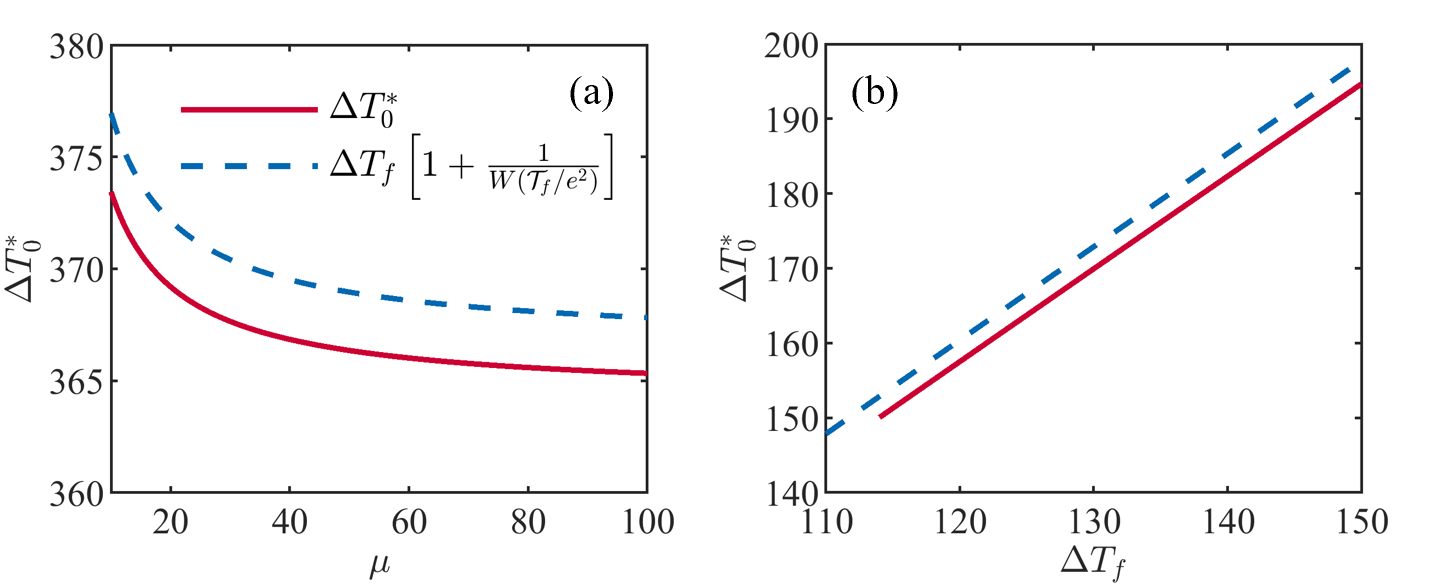}
    \caption{The behavior of the location of maximum cooling time in the regime of $\mathcal{M}\Delta T_f\gg\gamma_0/C$. The red solid lines illustrates the numerical solution of the extremum condition Eq.~\eqref{eq:maximum condition},  and the blue dashed lines denote Eq.~\eqref{eq:T0*-large-0}. (a) The dependence of $\Delta T_0^*$ on $\mu$ with $\Delta T_f=300$ and $\Theta=1$. (b) The dependence of $\Delta T_0^*$ on $\Delta T_f$ with $\mu=5$ and $\Theta=0.8$. Other parameters are fixed at $C=1$, $n_0=0.1$ and $\Gamma_2=5$.}
    \label{fig:T0*-1}
\end{figure}

Consider the opposite limit, $\mathcal{M}\Delta T_f \gg \gamma_0/C$, i.e., $  \mathcal{T}_f \gg 1$, where the maximum point is expected to approach the target temperature difference. We therefore introduce the relative deviation
\begin{equation}
    \delta
    =
    \frac{\Delta T_0^*-\Delta T_f}{\Delta T_f}
    =
    \frac{\mathcal{T}_0-\mathcal{T}_f}{\mathcal{T}_f},
    \label{eq:delta T}
\end{equation}
thus $\mathcal{T}_0= \mathcal{T}_f(1+\delta)$. Substituting Eq.~\eqref{eq:delta T} into Eq.~\eqref{eq:maximum condition}, we obtain
\begin{equation}
   2+\frac{1}{\mathcal{T}_f(1+\delta)}+\frac{\mathcal{T}_f}{1+\delta \mathcal{T}_f}
    =
    \ln\left[
    (1+\delta)(1+\delta\mathcal{T}_f)
    \right].
    \label{eq:maximumT0-1}
\end{equation}
If $\delta=O(1)$, the left-hand side remains finite, whereas the right-hand side contains a large contribution of order $\ln\mathcal{T}_f$. On the other hand, if $\delta\mathcal{T}_f=O(1)$, the left-hand side is of order $\mathcal{T}_f$, while the right-hand side is finite.  Therefore, the self-consistent scaling regime of $\delta$ is
\begin{equation}
    \frac{1}{\mathcal{T}_f}\ll\delta\ll1 .
    \label{eq:sacling delta}
\end{equation}
With Eq.~\eqref{eq:sacling delta}, Eq.~\eqref{eq:maximumT0-1} reduces to
\begin{equation}
    \frac{1}{\delta}+2
    \simeq
    \ln(\delta\mathcal{T}_f).
    \label{eq:maximu condition-large}
\end{equation}
yielding the maximum point 
\begin{equation}
    \Delta T_0^*
    \simeq
    \Delta T_f
    \left[
    1+
    \frac{1}{
    W\left(
    \dfrac{\mathcal{M}C\Delta T_f}{e^2\gamma_0}
    \right)
    }
    \right] \sim \Delta T_f+
    \frac{\Delta T_f}{
    \ln\left(
    \dfrac{\mathcal{M}C\Delta T_f}{\gamma_0}
    \right)
    }.
    \label{eq:T0*-large-0}
\end{equation}
Eq.~\eqref{eq:T0*-large-0} indicates that, in the regime $\mathcal{M}\Delta T_f\gg \gamma_0/C$, $\Delta T_0^*$ exhibits the same qualitative dependence on $\mathcal{M}$ and $\Delta T_f$ as in the regime $\mathcal{M}\Delta T_f\ll \gamma_0/C$. In addition, as $\mathcal{M}$ increases, the maximum point approaches $\Delta T_f$ from above. Specifically, only the excess part $\Delta T_0^*-\Delta T_f$ decreases with increasing $\mathcal{M}$, with the decrease being logarithmically slow, implying that the phase boundary becomes much flatter at large $\mathcal{M}$. Fig.~\ref{fig:T0*-1} shows the dependence of $\Delta T_0^*$ on $\mu$ in the regime $\mathcal{M}\Delta T_f\gg\gamma_0/C$. The direct numerical solution agrees well with the asymptotic approximation, with only a slight deviation. Moreover, as shown in Fig.~\ref{fig:T0*-1}(b), $\Delta T_0^*$ depends approximately linearly on $\Delta T_f$, with a logarithmic correction that keeps the maximum point above $\Delta T_f$.

\section{Supplementary Analysis of Example II}
\label{Example II}
Here we provide the detailed derivation and the corresponding diagrams for Example II given in the main text. The evolution of temperature is governed by [Eq.~(6)] with $\mathcal{I}=0$, i.e.,
\begin{equation}
    \frac{\mathrm{d}T}{\mathrm{d}t}=-\left[\frac{\Lambda_1-\alpha_0(T_0)(\Lambda_1-\Lambda_2)}{\tilde{C}}+\kappa(\Lambda_1-\Lambda_2)(T_0-T)\right](T-T_r).
    \label{eq:temperature evolution-2}
\end{equation}
Solving Eq.~\eqref{eq:temperature evolution-2}, we obtain the temperature difference and the cooling time, which take the identical expression of [Eq.~(11)] and [Eq.~(12)], respectively. It is worth noting that $\gamma_0$ is a constant in Example I, however, it is now a function of initial temperature $T_0$, namely, $\gamma_0=\gamma_0(T_0)$. Therefore, the emergence of ME is not only determined by the kinetic response coefficient $\mathcal{M}$ but also by $\gamma_0(T_0)$, characterized by $\partial\gamma_0/\partial T_0$. The criterion for anomalous relaxation presented in the main text reads
\begin{equation}
    \frac{\partial \gamma }{\partial T_0} = \frac{\Lambda_1 - \Lambda_2}{\tilde{C}} \left[ \kappa \tilde{C} - \frac{\partial \alpha_0}{\partial T_0} \right].   \label{eq:gamma_derivative}
\end{equation}
A positive correlation between $\gamma$ and $T_0~(\partial\gamma/\partial T_0>0)$  induces the ME during cooling process, since a hotter initial state leads to a higher effective relaxation rate. And a negative correlation $(\partial\gamma/\partial T_0<0)$ triggers the Inverse ME in heating process. 

In the simplest situation where $\alpha_0$ is a constant, Eq.~\eqref{eq:gamma_derivative} is solely depends on the sign of $\mathcal{M}=\lambda \Lambda_1\kappa/\tilde{C}$, where we have introduced $\lambda\equiv1-\Lambda_2/\Lambda_1$. Here, $\mathcal{M}>0$ accelerates cooling, whereas $\mathcal{M} < 0$ accelerates heating by dynamically enhancing the effective heating rate, as the system absorbs heat ($Q<0$).

\begin{figure}
    \centering
    \includegraphics[width=\linewidth]{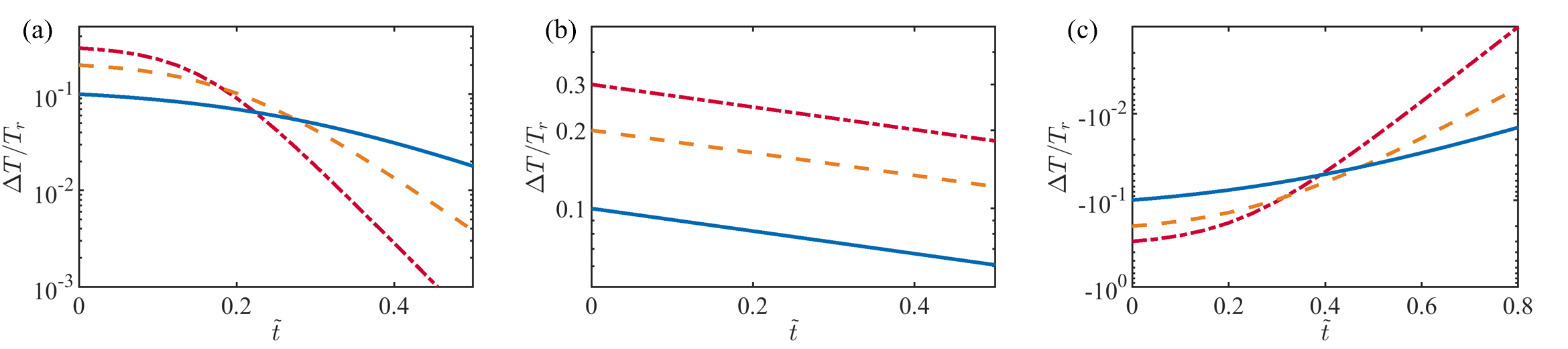}
    \caption{Dimensionless temperature difference $\Delta T/T_r$ as a function of dimensionless time $\tilde{t}=\gamma_0t$ with $\lambda=0.5$ (a), $\lambda=0$ (b) and $\lambda=-0.5$ (c). With $\Lambda_1=10$ and $\gamma_0=2.5$, the red dash-dotted curve, orange dashed curve, and blue solid curve in (a-b) are plotted with $\Delta T_0/T_r=0.3, 0.2,0.1$, respectively.  While the red dash-dotted curve, orange dashed curve, and blue solid curve in (c) with $\Lambda_1=5$ and $\gamma_0=5$ are plotted with $\Delta T_0/T_r=-0.3$, $\Delta T_0/T_r=-0.2$, and $\Delta T_0/T_r=-0.1$. Other parameters are fixed at $\alpha_0=1$, $\kappa\Omega=0.5$, $\tilde{C}T_r/\Omega=120$ and $C=1$.}
    \label{fig:T-t_const}
\end{figure}

Figure~\ref{fig:T-t_const}(a) plots the temperature evolution against dimensionless cooling time $\tilde{t}=\gamma_0t$ with $\lambda=0.5$. The crossover of the trajectories with different initial temperatures indicates the existence of the ME. For comparison, with $\lambda=0$ (i.e. $\mathcal{M}=0$), Fig.~\ref{fig:T-t_const}(b) illustrates the normal cooling process, where the curves are parallel on a logarithmic scale. In addition, $\lambda<0$ triggers the Inverse ME, a system with a lower initial temperature heats up faster than a warmer one, as illustrated in Fig.~\ref{fig:T-t_const}(c) with $\lambda=-0.5$.

Consider the general situation where the initial structural variable depends on the initial temperature, $\alpha_0=\alpha_0(T_0)$. In this case, $\gamma_0$ becomes an explicit function of $T_0$; thus, the dependence of the relaxation rate on initial temperature is no longer determined solely by $\mathcal{M}$, but by the structural feedback during relaxation and the dependence of the initial structural variable on the initial temperature. Equation~\eqref{eq:gamma_derivative} captures the  initial-state dependence of the relaxation rate through two contributions: conductance contrast $\Lambda_1-\Lambda_2$ and structural sensitivity $\partial\alpha_0/\partial T_0$. With the reduced structural sensitivity 
\begin{equation}
    a\equiv\frac{\partial\alpha_0/\partial T_0}{\kappa\tilde{C}},
\end{equation}
the sign of $\partial\gamma/\partial T_0$ is determined by $\lambda(1-a)$. If $\lambda$ and $(1-a)$ share the same sign ($\partial\gamma/\partial T_0>0$), a hotter system possesses a larger relaxation rate, triggering the ME. Conversely, if they have opposite signs, the inverse ME can occur. The previous case with $\alpha_0=\mathrm{const}$ corresponds to the special case of $a=0$. 

\begin{figure}
    \centering
    \includegraphics[width=\linewidth]{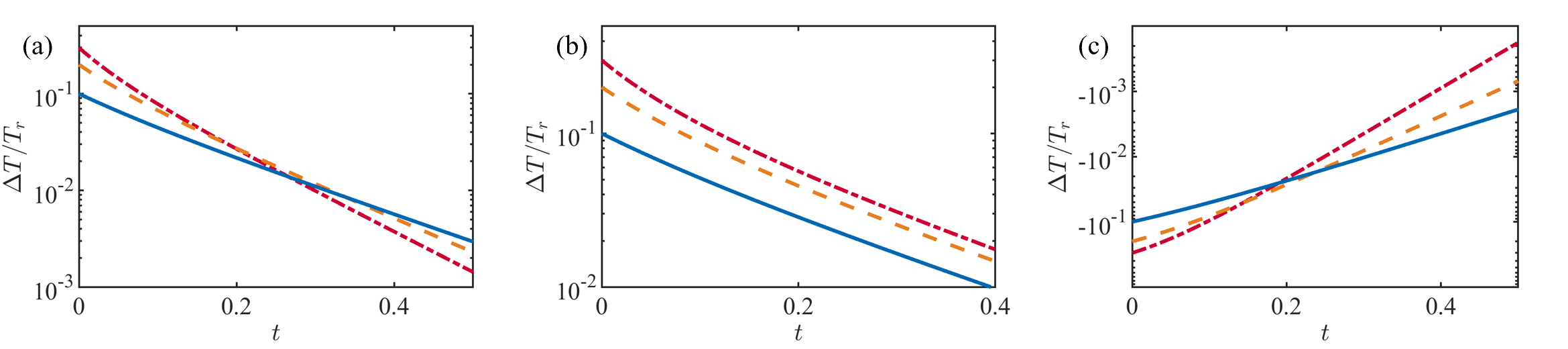}
    \caption{Dimensionless temperature difference $\Delta T/T_r$ as a function of time $t$ with $a=1.5$ (a), $a=1$ (b) and $a=-0.5$ (c). The red dash-dotted curve, orange dashed curve, and blue solid curve in (a-b) are plotted with $\Delta T_0/T_r=0.3, 0.2,0.1$, respectively.  While the red dash-dotted curve, orange dashed curve, and blue solid curve in (c) are plotted with $\Delta T_0/T_r=-0.3$, $\Delta T_0/T_r=-0.2$, and $\Delta T_0/T_r=-0.1$. Other parameters are fixed at $\lambda=-3$, $\kappa\Omega=0.028$, $\tilde{C}T_r/\Omega=61.7$, $\Lambda_1=5$, $C=1$ and $T_r=300$.}
    \label{fig:T-t_aT0}
\end{figure}

In the linear response regime for the initial structural variable on the initial temperature,
\begin{equation}
    \alpha_0=a\kappa\tilde{C}(T_0-T_r).
    \label{eq:alpha0}
\end{equation}
It follows from Eqs.~\eqref{eq:temperature evolution-2} and \eqref{eq:alpha0} that the evolution of temperature is solved, as illustrated in Fig.~\ref{fig:T-t_aT0} with $\lambda<0$ ($\mathcal{M}<0$). During the cooling process, the structural response hinders heat transfer, i.e., $\gamma_0+\mathcal{M}Q$ decreases with time, manifested by the decreasing magnitude of the slope of the cooling curves in Fig.~\ref{fig:T-t_aT0}(a-b). However, the dependence of $\gamma_0$ on the initial temperature reverses the initial-state dependence of the effective relaxation rate, leading to an overall positive correlation between the relaxation rate and the initial temperature. As shown in Fig.~\ref{fig:T-t_aT0}(a), a hotter system relaxes faster than a cooler one with $a=1.5$. For comparison, Fig.~\ref{fig:T-t_aT0}(b) illustrates the cooling process with $a=1$, where the relaxation rate does not depend on the initial temperature ($\partial\gamma/\partial T_0=0$). Notably, unlike the previous case, the cooling curves are no longer parallel on a logarithmic scale. This can be understood as follows: substituting Eq.~\eqref{eq:alpha0} with $a=1$ into the relaxation rate $\gamma=\gamma_0+\mathcal{M}Q$, we obtain
\begin{equation}
    \gamma=\frac{\Lambda_1}{\tilde{C}}-\kappa(\Lambda_1-\Lambda_2)T(t),
    \label{eq:r-T}
\end{equation}
changing the conventional linear Newton's cooling law to a nonlinear form ($\mathrm{d}T/\mathrm{d}t\propto-T(T-T_r)$). The red dash-dotted curve in Fig.~\ref{fig:T-t_aT0}(b) is steeper than the other two curves at the early stage of cooling. Additionally, $a<1$ with $\lambda<0$ induces the inverse ME, as shown in Fig.~\ref{fig:T-t_aT0}(c).

\begin{figure}
    \centering
    \includegraphics[width=\linewidth]{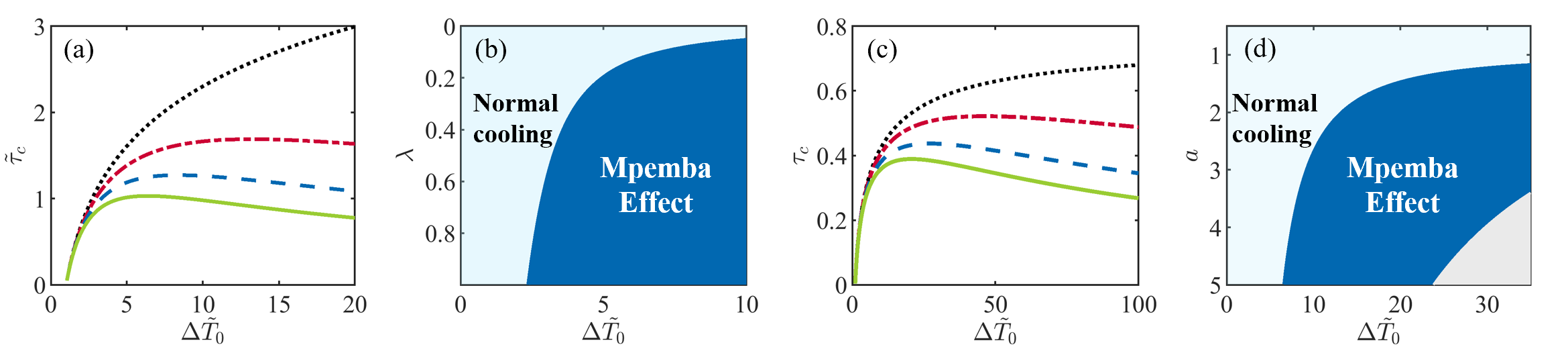}
    \caption{The occurrence of the ME within Example II. (a) With $\alpha_0=1$, the dimensionless cooling time $\tilde{\tau}_c=\gamma_0\tau_c$ as a function of dimensionless initial temperature difference $\Delta\tilde{T_0}=\Delta T_0/\Delta T_f$ with $\lambda=0, 0.1, 0.2, 0.3$ (from top to bottom). Other parameters are fixed at $\kappa\Omega=0.5$, $\tilde{C}\Delta T_f/\Omega=2.4$, $\Lambda_1=5$, $\Delta T_f=6$ and $C=1$. (b) Phase diagram of cooling in the parameter space $(\Delta\tilde{T}_0,\lambda)$. (c) With $\alpha_0=a\kappa\tilde{C}\Delta T_0$, the exact cooling time $\tau_c$ as a function of dimensionless initial temperature difference $\Delta\tilde{T_0}=\Delta T_0/\Delta T_f$ with $a=1, 1.2, 1.5,1.8$ (from top to bottom). Other parameters are fixed at $\lambda=-5$, $\kappa\Omega=0.0085$, $\tilde{C}\Delta T_f/\Omega=0.6$, $\Lambda_1=5$, $\Delta T_f=3$ and $C=1$. (d) Phase diagram of cooling in the parameter space $(\Delta\tilde{T}_0,a)$. The gray region denotes the inadmissible regime, where $\alpha_0\in[0,1]$ is violated.}
    \label{fig:tau_phase}
\end{figure}

Furthermore, Fig.~\ref{fig:tau_phase}(a,c) shows the cooling time required to reach a target final temperature difference $\Delta T_f$ for $\alpha_0=\mathrm{const}$ and $\alpha_0=a\kappa\tilde{C}\Delta T_0$, respectively. The black dotted curves denote the memoryless limit ($\partial \gamma/\partial T_0=0$), in which the cooling time monotonically increases with $\Delta \tilde{T}_0$. Larger $\partial \gamma/\partial T_0$ induces a
progressively steeper drop in cooling time. The occurrence of the ME in Example II is also characterized by the order parameter $\chi$ defined in the main text. The phase diagram plotted in Fig.~\ref{fig:tau_phase}(b, d) corresponds to  $\alpha_0=\mathrm{const}$ and $\alpha_0=a\kappa\tilde{C}\Delta T_0$, respectively. In particular, the gray region in Fig.~\ref{fig:tau_phase}(d) denotes the inadmissible parameter regime where $\alpha_0>1$. 

\bibliography{Refs}